\documentclass{PoS}

\renewcommand{\Re}{\rm Re}

\title{LQCD: Flavor Physics and Spectroscopy}

\ShortTitle{LQCD: Flavour Physics and Spectroscopy}

\author{\speaker{Carleton DeTar}\\
        Department of Physics and Astronomy, University of Utah, Utah 84112, USA\\
        E-mail: \email{detar@physics.utah.edu}}

\abstract{I review highlights of recent results in quark-flavor physics and
  heavy-quark spectroscopy from lattice QCD, with emphasis on
  $B$-meson decays and excited and exotic charmonium states. }

\FullConference{International Symposium on Lepton Photon Interactions at High Energies\\
		17-22 August 2015\\
		University of Ljubljana, Slovenia}

\begin{document}

\section{Introduction}

The numerical simulation of lattice quantum chromodynamics (QCD) has
made enormous strides over the past two decades thanks to advances in
computing power and algorithms.  It is now the method of choice for
studying QCD in the static, nonperturbative regime, {\it e.g.} for
such properties as hadronic structure, spectroscopy, and transition
amplitudes and for studying hot and dense QCD in thermal equilibrium.
The lattice serves as an ultraviolet regulator, and, with suitable
renormalization and the continuum limit, it is expected to fall in the
same universality class as QCD with any of the popular continuum
regulators.  Thus lattice QCD is an {\it ab-initio} method in that its
results can be refined to arbitrary precision (given enough computing
power).  There are no uncontrolled model approximations.

\begin{figure}
   \centering
   \includegraphics[width=0.5\textwidth]{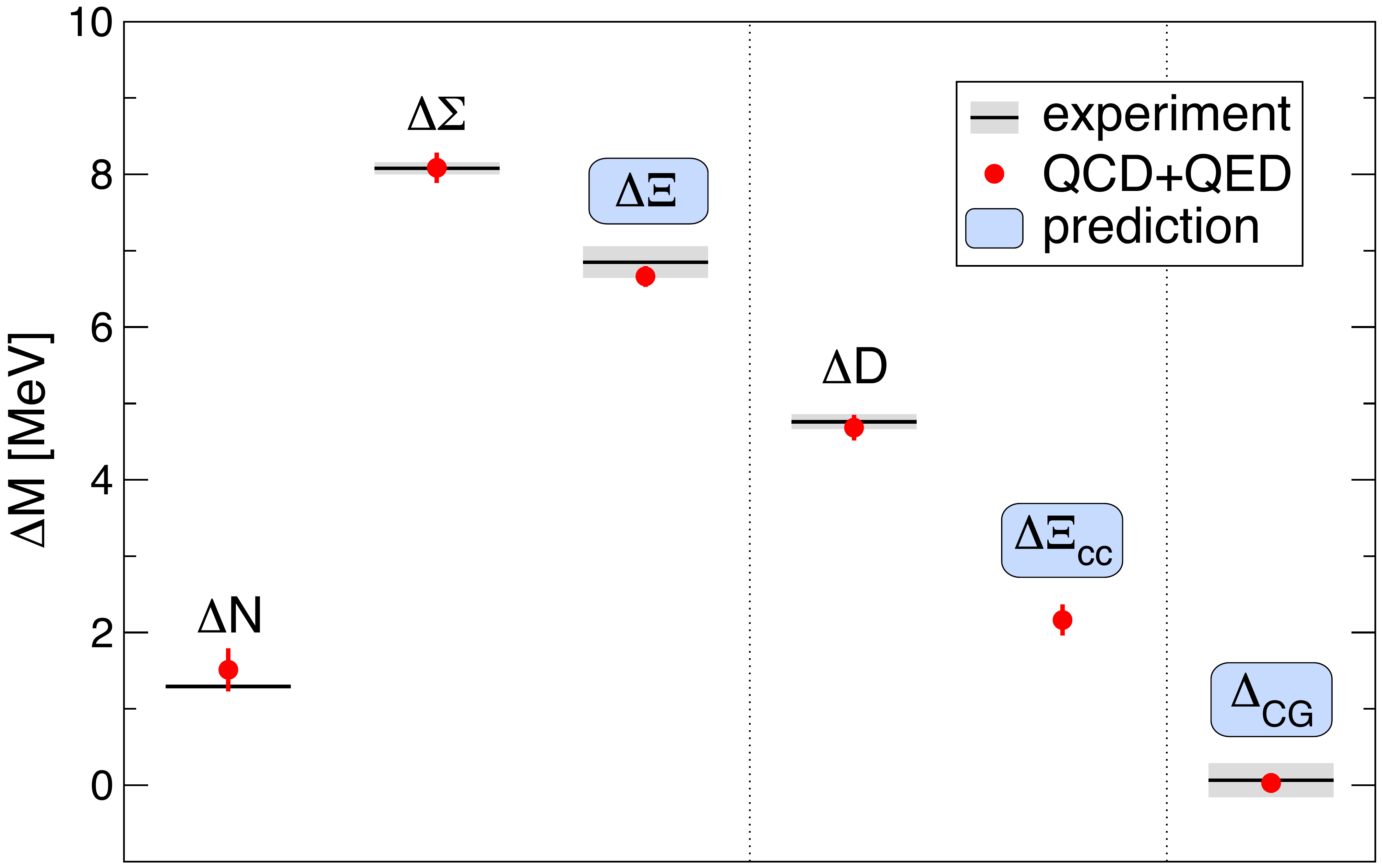}
  \caption{Charge-multiplet mass splittings of the $\frac{1}{2}^+$
    baryon octet from a lattice QED+QCD calculation compared
    with the experimental values \cite{Borsanyi:2014jba}.
    Such calculations are excellent tests of the lattice methodology. }
 \label{fig:octetEM}
\end{figure}

We validate our numerical methodology by comparing results of
calculations with well-established measurements.  An impressive recent
example is the Budapest-Marseille-Wuppertal calculation of the
electromagnetic splittings of the baryon octet, shown in
Fig.~\ref{fig:octetEM} \cite{Borsanyi:2014jba}.  This lattice
calculation included both QCD and quantum electrodynamics.

A variety of lattice formulations of QCD are in wide current use.
They differ particularly in their fermion formulation.  Each has its
good and bad points.  However, all are expected to result in the same
theory in the continuum limit.  Thus it is useful to carry out
important calculations with more than one lattice formulation as a
cross check.  For quark-flavor physics, the Flavor Lattice Averaging Group
provides biennial reviews of recent lattice results drawing from the
wide variety of lattice formulations \cite{Aoki:2013ldr}.  A new
review is expected later in 2015.

In assessing the quality of the result of a lattice calculation, one
should look for the following features: (1) Was the calculation done
at multiple lattice spacings and has the continuum limit been taken?
(2) Was the lattice volume large enough that finite volume effects are
under control? (3) What sea quark flavors have been included?  Many
studies now include charm sea quarks as well as strange, up, and down,
but some still include only up and down.  (4) Were all quark masses at
their physical values? If not, was a suitable
extrapolation/interpolation carried out to reach the physical masses?
(5) Was a complete analysis of systematic errors undertaken?

To be sure, lattice QCD has its limitations. It is not well suited to
the study of real-time behavior.  Phenomena such as the evolution of
hadronic jets and multiparticle scattering are not easily treated.  At
high temperature, only thermodynamic equilibrium and perturbations
thereof, including transport properties, are feasible \cite{UkawaLP}.
Out-of-equilibrium processes are very difficult subjects.

My task in this brief talk is to review recent lattice-QCD highlights
in quark-flavor physics and in spectroscopy.  For quark-flavor
physics, I will focus on $B$-meson decays and tests of the standard
model.  For spectroscopy, I will describe progress in studies of
excited and exotic charmonium states.

\section{Quark-flavor-physics highlights}

The experimental and theoretical flavor-physics programs aim to obtain
accurate values of standard-model parameters and especially to subject
the standard model to stringent tests in the hope of discovering
evidence for new physics.  High-precision tests extend the high-energy
reach of experiment.  Tree-level leptonic and semileptonic decays of
heavy mesons to light charged leptons are thought to provide reliable
determinations of standard-model parameters free of significant
new-physics contamination.  Rare decays of heavy mesons involving
higher-order electroweak processes are hoped to be a particularly
promising place to find evidence for physics beyond the standard
model.  The same is true of neutral-meson mixing, which is
intrinsically higher order. Although these processes arise from
electroweak interactions, any process involving hadrons necessarily
also involves QCD, so lattice QCD is needed in order to simulate the
decay environment and expose the underlying electroweak physics.

\begin{figure}
  \centering
   \includegraphics[width=0.3\textwidth]{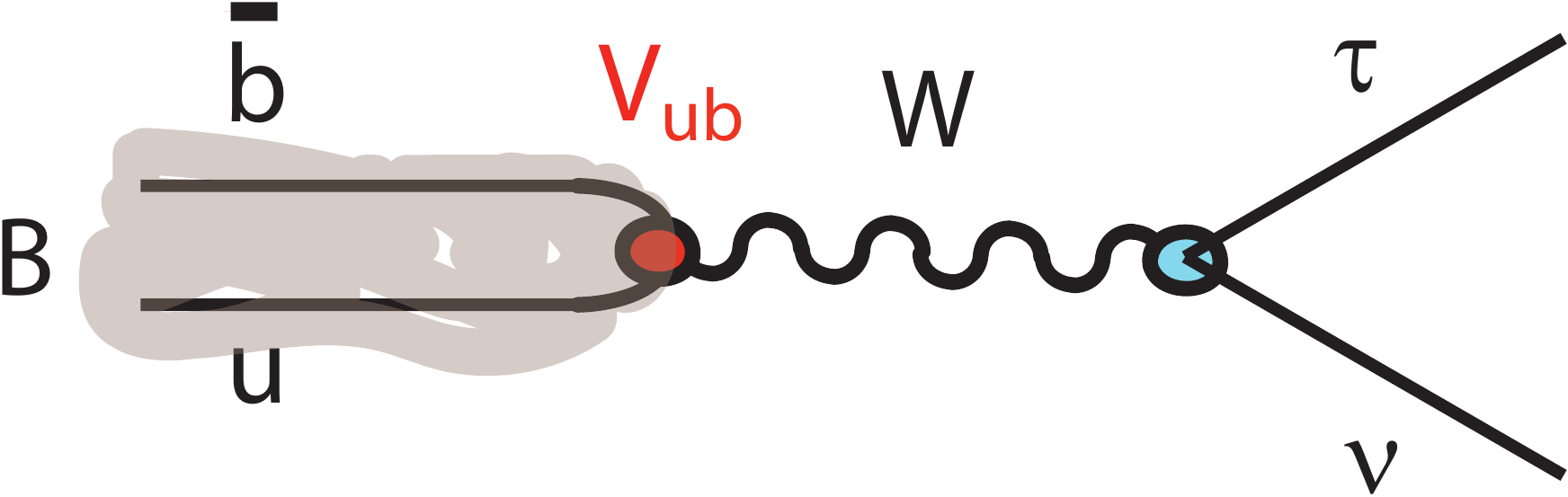} \hfill
   \includegraphics[width=0.3\textwidth]{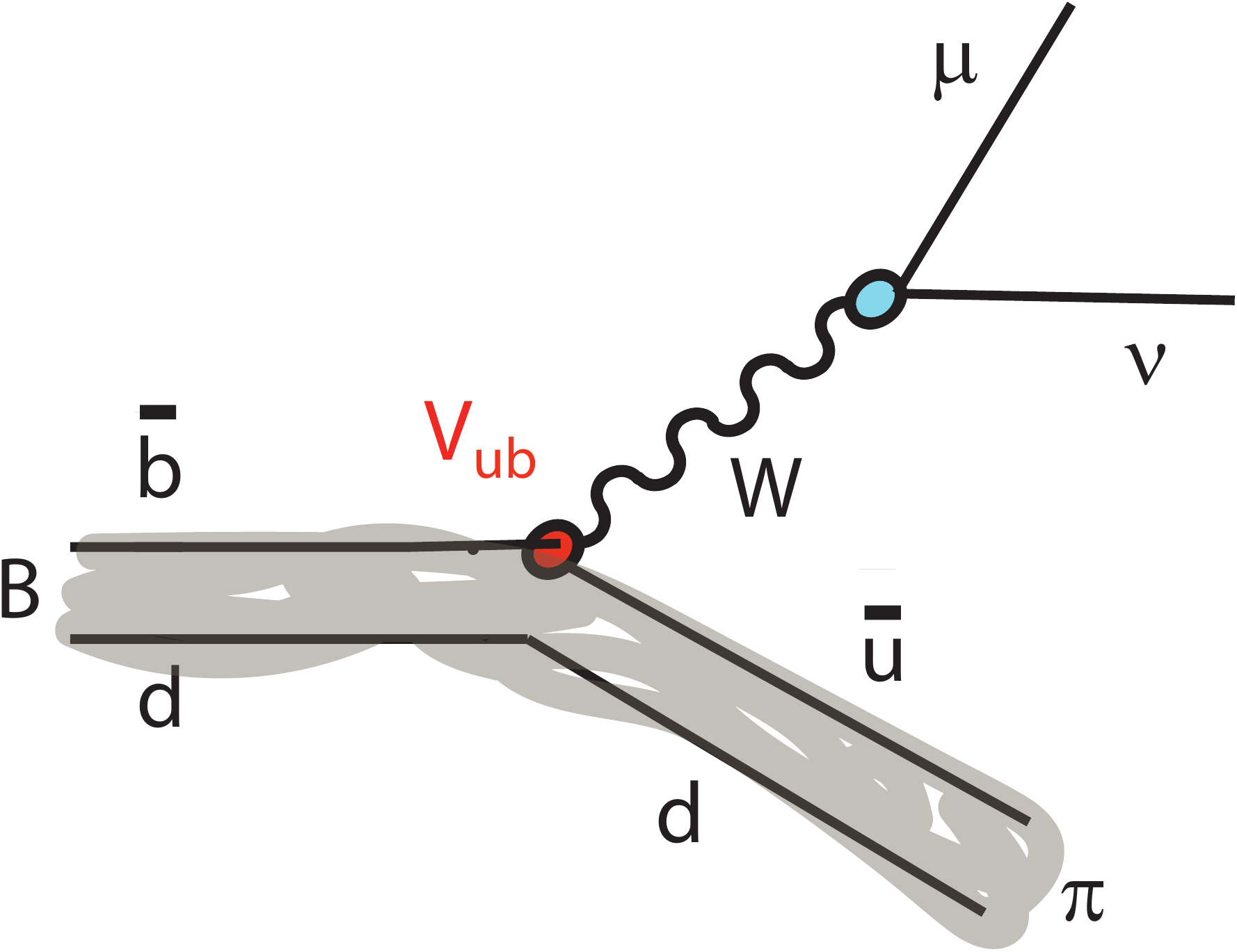} \hfill
   \includegraphics[width=0.3\textwidth]{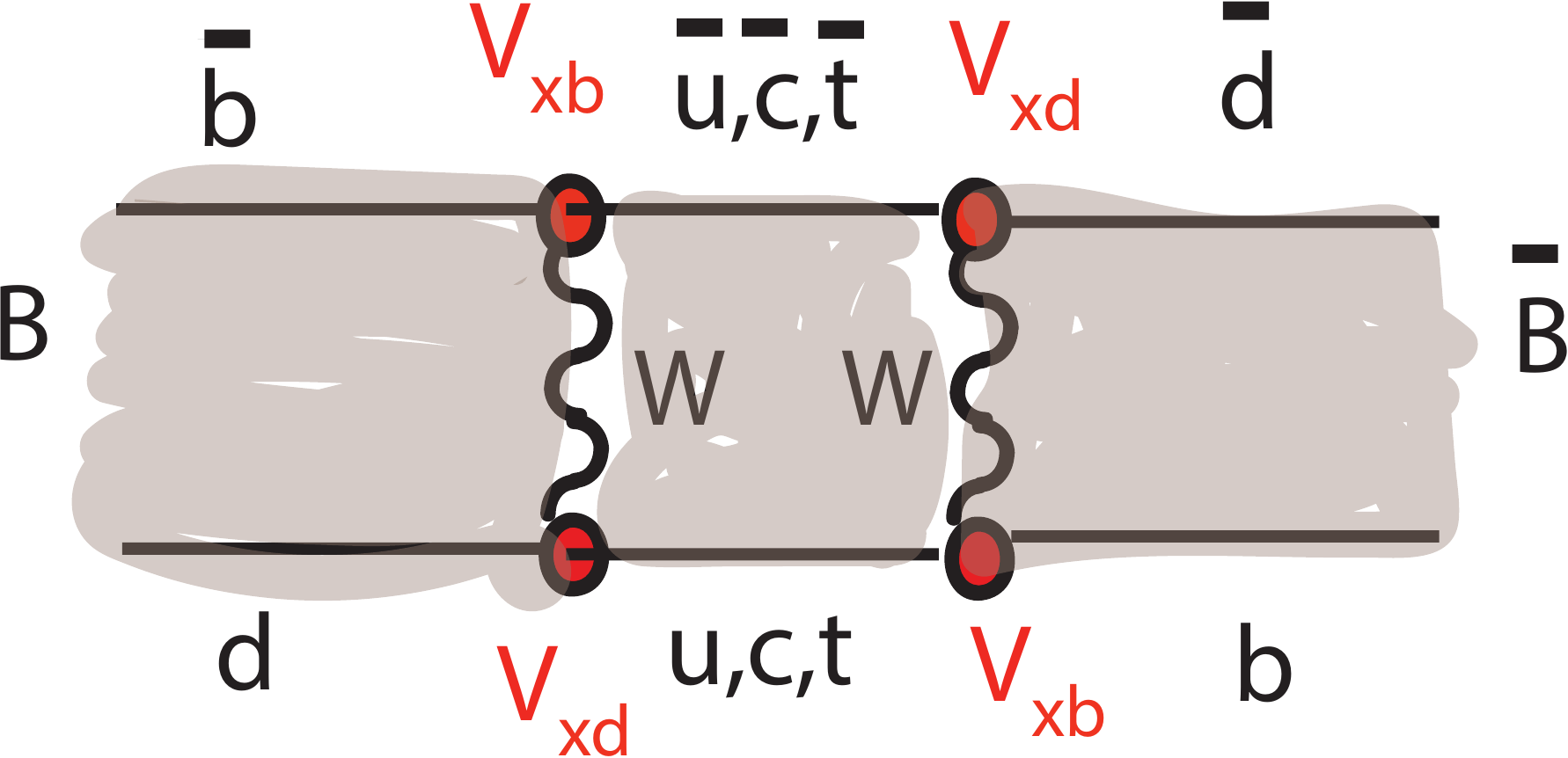}
  \caption{Typical valence quark-line diagrams for $B$-meson
    processes: (left) leptonic decay, (middle) semileptonic decay, and
    (right) one of the neutral $B$-meson mixing diagrams. The shaded
    regions denote strong interactions.}
 \label{fig:diagrams}
\end{figure}

The Cabibbo-Kobayashi-Maskawa (CKM) matrix parameterizes the mixing
between quark flavors under charged weak interactions.  In the
standard model it is unitary.  Thus a popular test checks unitarity.
CKM matrix elements are typically determined from flavor transitions
in leptonic or semileptonic decays or in neutral meson mixing.  Some
relevant valence quark-line diagrams are shown in
Fig.~\ref{fig:diagrams}.  The underlying electroweak processes are
treated in perturbation theory, but the strong interactions are
treated nonperturbatively in lattice gauge theory.  So, for example,
the standard model $B \to \pi \ell \nu$ differential decay rate
\begin{equation}
   d\Gamma/dq^2 \propto |V_{ub}|^2 |f_+(q^2)|^2 \, ,
\end{equation}
depends on the CKM matrix element $|V_{ub}|$ and the nonperturbative
hadronic form factor $|f_+(q^2)|$, which is computed in lattice QCD.
The proportionality constant is a product of a known kinematic factor
and the squared Fermi constant.  So we can solve for $|V_{ub}|$ and
use the result in the test of CKM unitarity.  The error in the result
is a combination of experimental and theoretical errors.  Clearly, to
make progress in precision, it is important that both errors be
reduced in parallel.

Similar considerations apply for leptonic decays and neutral meson
mixing.  Recent summaries of results for CKM matrix elements may be
found in
Refs.~\cite{Aoki:2013ldr,Bouchard:2015pda,RuthCIPANP,PenaLat15,RuthEPS2015,Rosner:2015wva}.

\begin{figure}
\centering
   \includegraphics[width=0.45\textwidth]{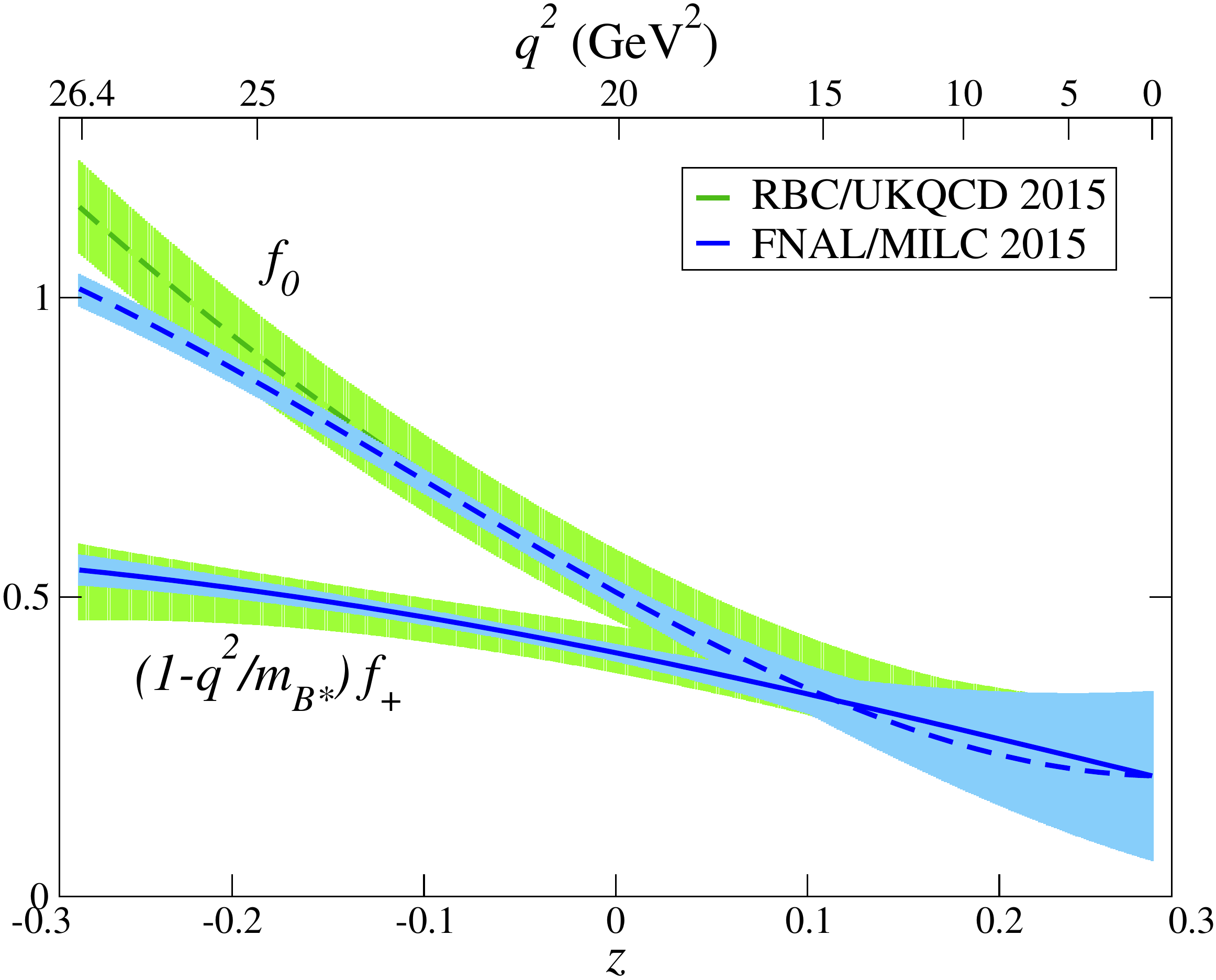} \hfill
   \includegraphics[width=0.45\textwidth]{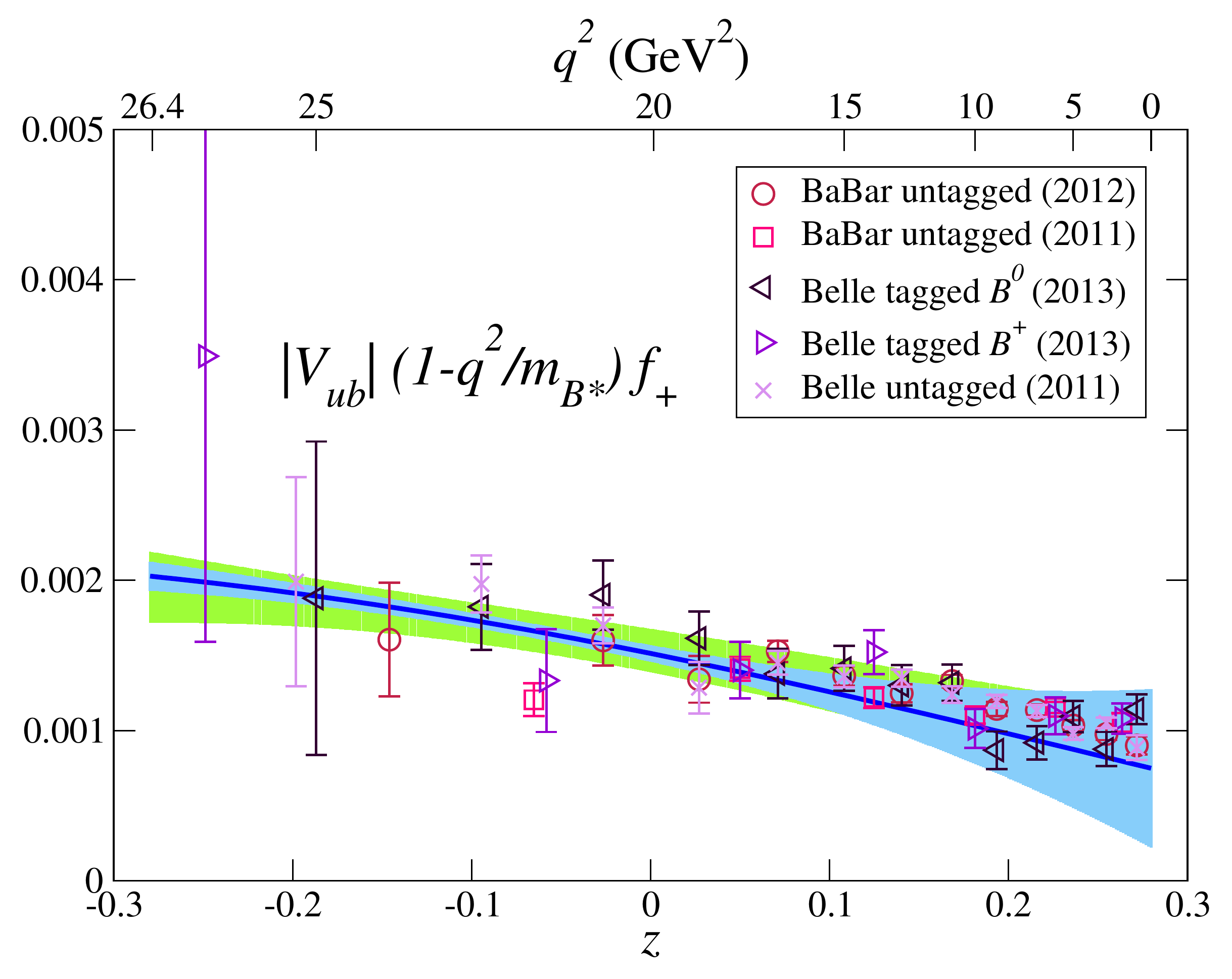}
  \caption{Left panel: a comparison of lattice results for the $B \to
    \pi$ vector ($f_+$) and scalar ($f_0$) form factors from the
    RBC/UKQCD collaboration (green band) \protect\cite{Flynn:2015mha}
    and from the Fermilab/MILC collaboration (blue band)
    \protect\cite{Lattice:2015tia}.  The lattice methods are different
    but the agreement is excellent. Right panel: a comparison of the
    theoretical form factors $f_+$ in the left plot with the
    experimental differential decay rate. The measured shape is nicely
    reproduced.  (Figure credit \cite{RuthEPS2015}).}
 \label{fig:Bpilnu}
\end{figure}

\subsection{$B \to \pi \ell \nu$ at nonzero recoil}

In this talk, I highlight mostly results involving $B$ physics for
which several new results were reported in the past few months.  A
principal challenge for $B$ physics is gaining control of the lattice
discretization errors.  This is achieved either through the Fermilab
approach \cite{ElKhadra:1996mp} or the RHQ variant thereof
\cite{Christ:2006us}, a nonrelativistic treatment
\cite{Lepage:1992tx}, or working with lattice spacings $a < 1/m_b
\approx 0.05$ fm (for $b$-quark mass $m_b$), followed by a careful
extrapolation to zero lattice spacing.  This has been done in two new
studies of the semileptonic decay $B \to \pi \ell \nu$.  Results for
the form factors from two different lattice methods are plotted in
Fig.~\ref{fig:Bpilnu}.  The RBC/UKQCD calculation \cite{Flynn:2015mha}
is based on ensembles generated with domain-wall quarks and uses the
RHQ action for the $b$ quark.  The Fermilab Lattice/MILC calculation
\cite{Lattice:2015tia} is based on ensembles generated with the asqtad
action with clover $b$ quarks in the Fermilab interpretation. It
yields an error of approximately 3.5\% in the vector form factor $f_+$
in the central kinematic range. The results are extended over the
whole kinematic range using a model-independent ``$z$'' expansion
based on the known analytic structure of the form factor.  Using this
expansion and fitting the Fermilab/MILC form factors with a
combination of recent Babar~\cite{delAmoSanchez:2010af,Lees:2012vv}
and Belle \cite{Ha:2010rf,Sibidanov:2013rkk} data then results in the
value $|V_{ub}| = 3.72(16) \times 10^{-3}$.

\subsection{$B \to D \ell \nu$ at nonzero recoil}

\begin{figure}
\centering
   \includegraphics[width=0.45\textwidth]{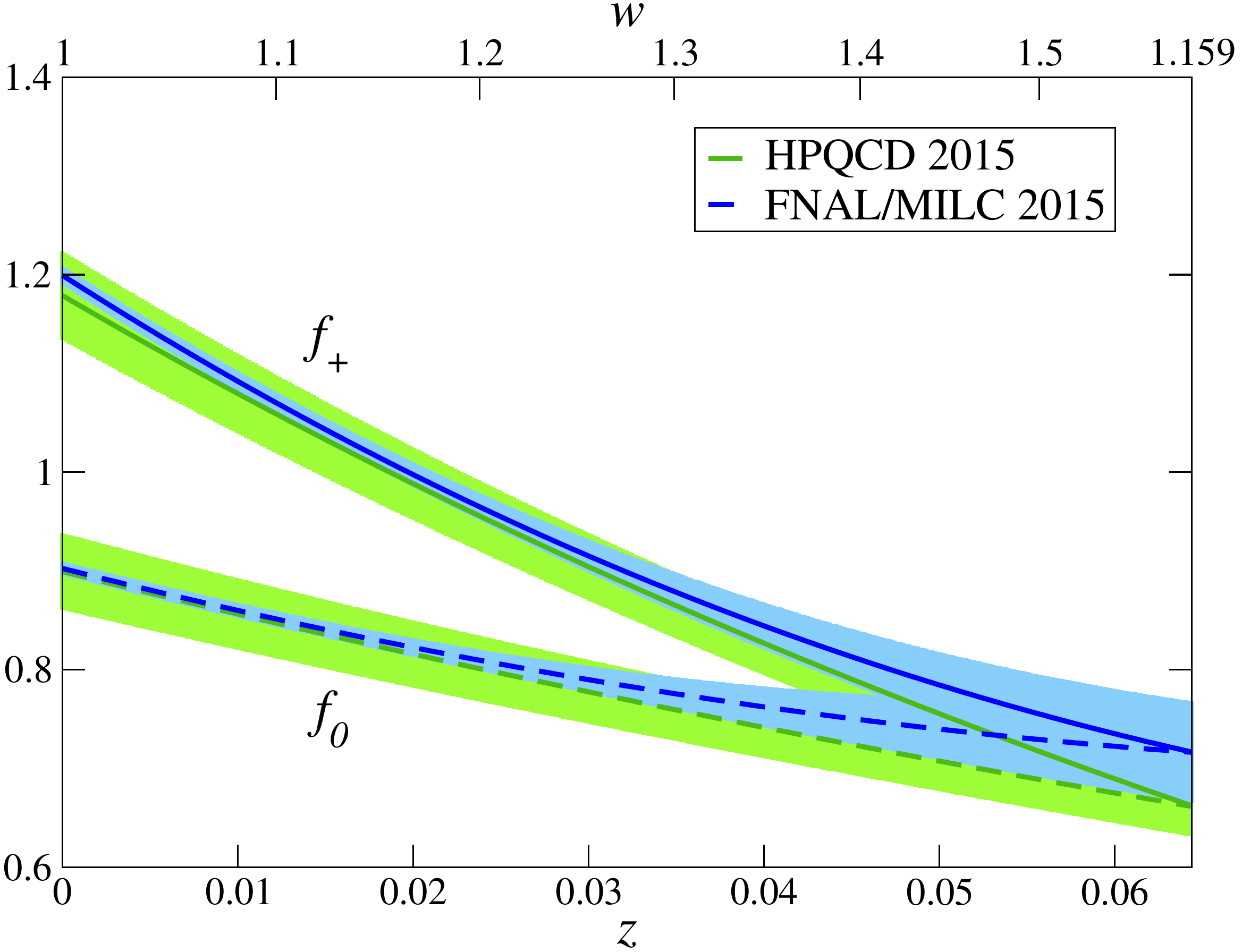} \hfill
   \includegraphics[width=0.5\textwidth]{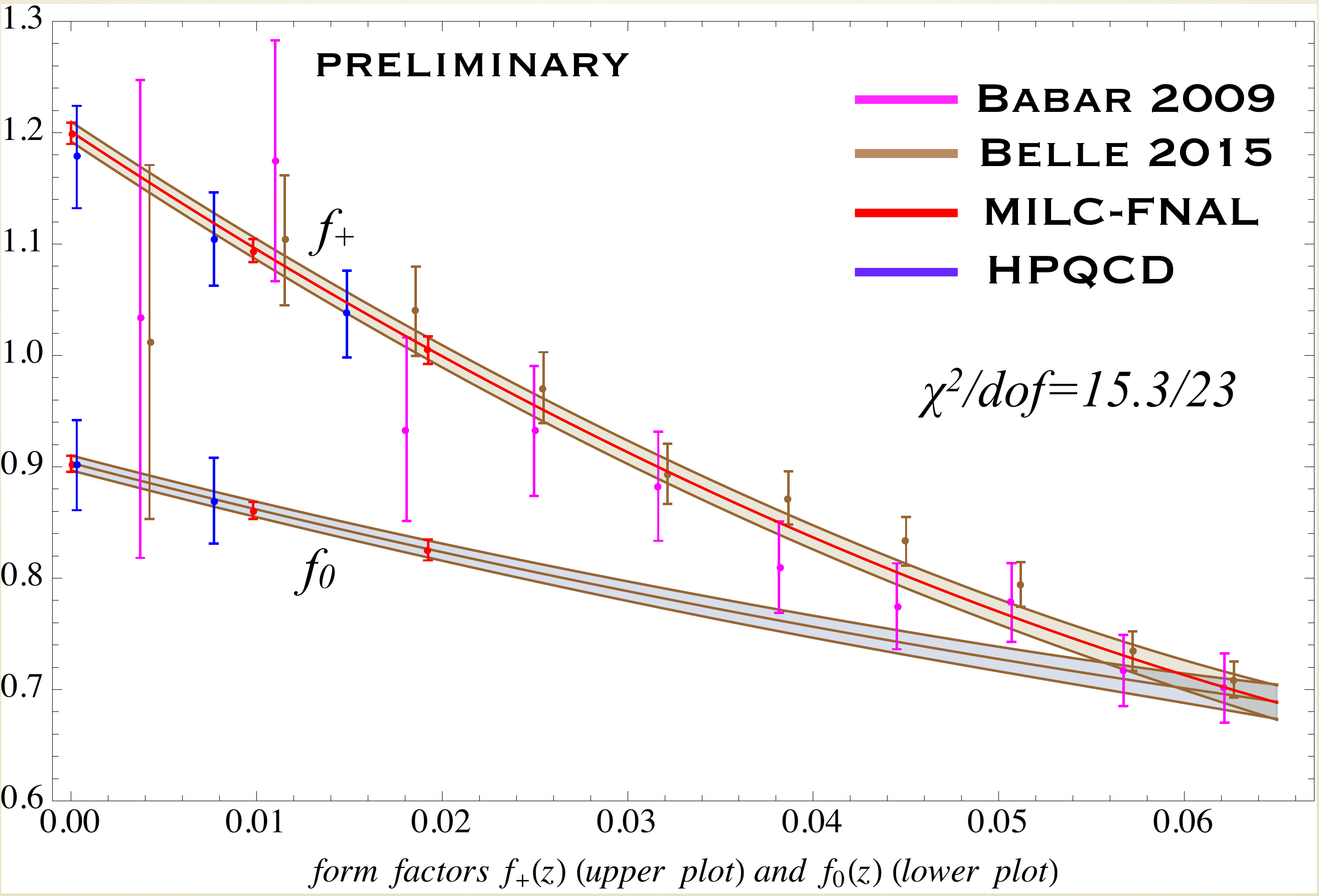}
  \caption{Left: Vector and scalar form factors for $B \to D \ell \nu$
    from two recent lattice calculations
    \protect\cite{Lattice:2015rga,Na:2015kha}.  The agreement is good.
    (Figure credit: \protect\cite{RuthCIPANP}).  Right: Joint fit of
    the form factors on the left and experimental data from
    BaBar \protect\cite{Aubert:2008yv} and preliminary data from Belle
    \protect\cite{GlattauerEPS2015,GambinoEPS2015}. (Figure credit:
    \cite{GambinoEPS2015}).  }
 \label{fig:B2D}
\end{figure}

A determination of $|V_{cb}|$ from the semileptonic decay $B \to D
\ell \nu$ rate is best done at nonzero recoil where the relative
experimental and theoretical uncertainties are small.  In the past
year, results from two new lattice calculations at nonzero recoil were
reported \cite{Lattice:2015rga,Na:2015kha}.  These were the first such
calculations to take into account the effects of sea quarks.  Both
groups used gauge-field ensembles created by the MILC collaboration
based on 2+1 flavors of asqtad sea quarks, resulting in some overlap
of inputs.  However, the treatments were otherwise independent.
Results for both the vector and scalar form factors are shown in the
left panel of Fig.~\ref{fig:B2D}.  The agreement is very good. A
combined fit including both new theoretical results and recent Babar
\cite{Aubert:2008yv} and Belle \cite{GlattauerEPS2015,GambinoEPS2015}
data is shown in the right panel, and yields the value $|V_{cb}| =
40.7(1.0)_{\rm latQCD+expt}(0.2)_{\rm QED}\times 10^{-3}$.

\subsection{$R(D)$}

Semileptonic decays of the $B$ meson, $B \to D \ell \nu$ and $B \to
D^* \ell \nu$ can result in a $\tau$ lepton as well as $e$ or $\mu$.
The quantities $R(D)$ and $R(D^*)$ are ratios of branching ratios for
decays producing a $\tau$ or a $\mu$:
\begin{equation}
  R(D^{(*)}) = \frac{\Gamma(B \to D^{(*)} \tau \nu)}{\Gamma(B \to D^{(*)}\mu \nu)} \, ,
\end{equation}
and are studied as a test for new physics.  In 2012 the BaBar
collaboration reported a value $R(D) = 0.440(58)(42)$ and $R(D^*) =
0.332(24)(18)$ \cite{Lees:2012xj}, which created some excitement
because it disagreed with a standard model expectation: $R(D) =
0.297(17)$ and $R(D^*) = 0.252(3)$ \cite{Kamenik:2008tj} at the
$3.5\sigma$ level. This year's new experimental results from the
Belle and LHCb collaborations, $R(D) = 0.375(64)(26)$ and $R(D^*) =
0.293(38)(15)$ \cite{Huschle:2015rga} and $R(D^*) = 0.366(27)(30)$
\cite{Aaij:2015yra}, confirm the discrepancy with the standard model;
the HFAG combination of all measurements disagrees at
3.9$\sigma$~\cite{Amhis:2014hma}.

\subsection{Exclusive {\it vs.} inclusive $|V_{cb}|$ and $|V_{ub}|$.}

\begin{figure}
\centering
   \includegraphics[width=0.48\textwidth]{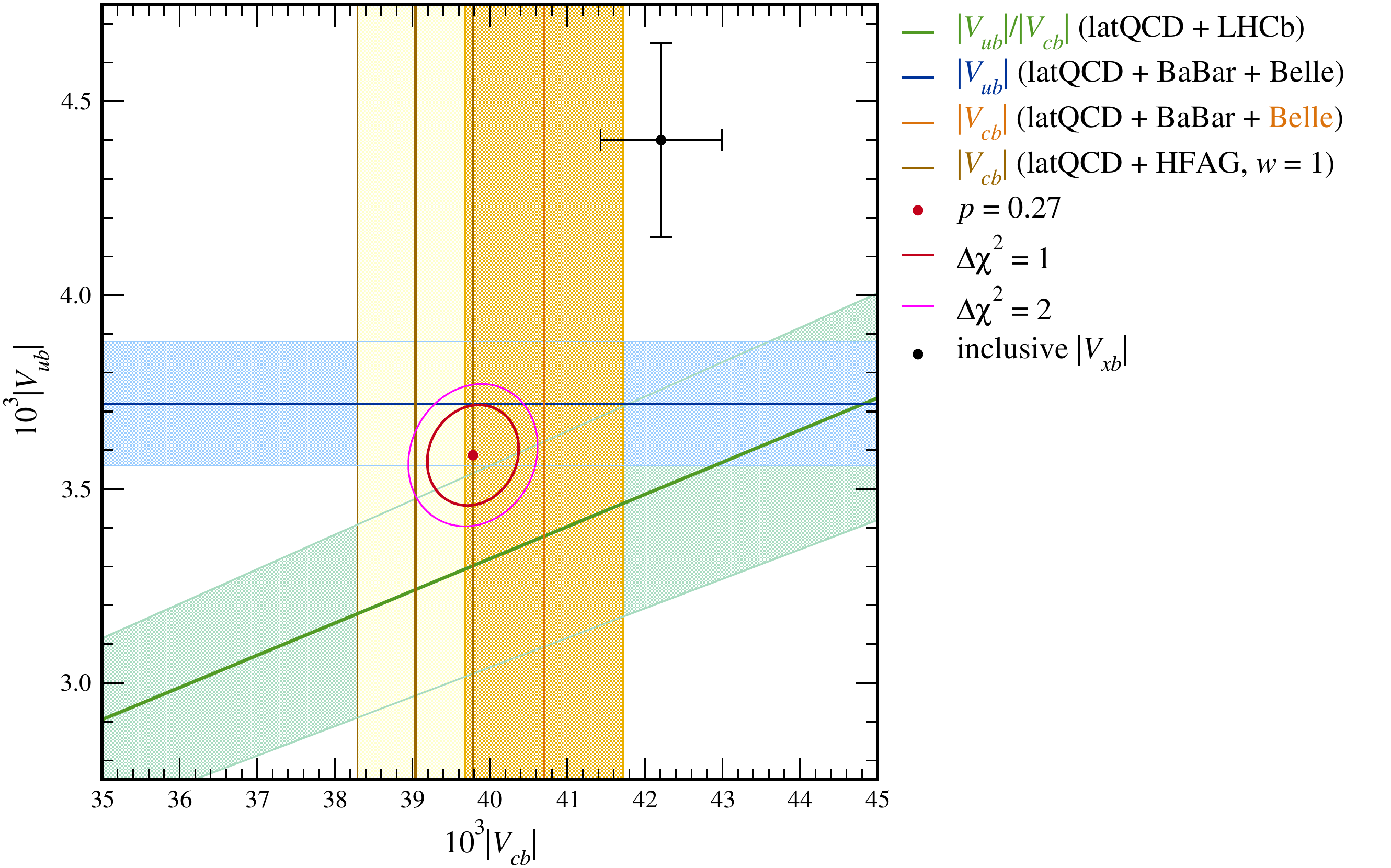} \hfill
   \includegraphics[width=0.5\textwidth]{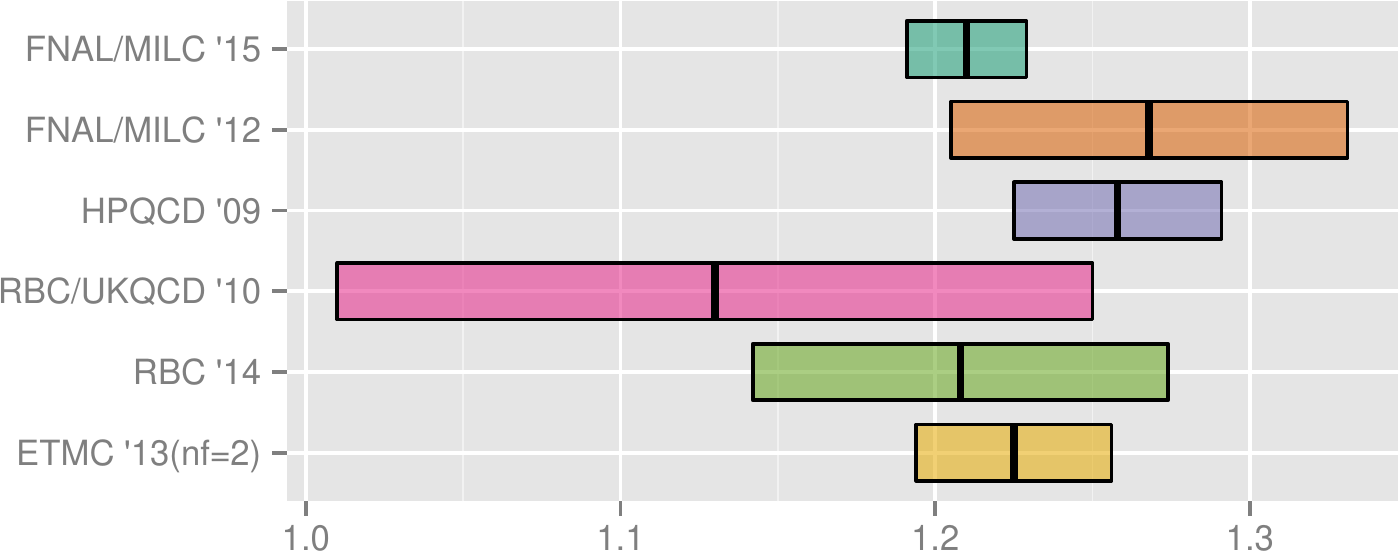}
  \caption{Left: determination of $|V_{cb}|$ and $|V_{ub}|$ using a
    fit to recent lattice results for the exclusive processes $B \to
    \pi\ell\nu$ \protect\cite{Flynn:2015mha,Lattice:2015tia} (blue
    band), $B \to D^*\ell\nu$ \protect\cite{Bailey:2014tva} (light
    yellow band), $B \to D\ell\nu$
    \protect\cite{Lattice:2015rga,Na:2015kha} (dark yellow band) and
    the ratio of differential decay rates $\Gamma(\Lambda_b \to p \ell
    \nu)/\Gamma(\Lambda_b \to \Lambda_c \ell \nu)$
    \protect\cite{Detmold:2015aaa} (pale green diagonal band).  The
    exclusive determinations are compatible at the $p = 0.27$
    level. The inclusive determinations
    \protect\cite{Amhis:2014hma,Alberti:2014yda}, shown by the cross, display a
    strong tension with the 1$\sigma$ and 2$\sigma$ error ellipses.
    (Figure credit: \cite{KronfeldVubVcb}). Right: comparison of a
    recent preliminary result for the $B$-mixing ratio $\xi$ from
    Ref.~\cite{SimoneLat15} with previous results
    \protect\cite{Bazavov:2012zs,Albertus:2010nm,Gamiz:2009ku,Carrasco:2013zta,Aoki:2014nga}.
  }
 \label{fig:excl_incl_xi}
\end{figure}

There has been a long-standing tension between determinations of
$|V_{cb}|$ and $|V_{ub}|$ from inclusive and exclusive $B$-meson
decays.  Figure~\ref{fig:excl_incl_xi} illustrates the severity of
this tension by plotting results from (1) the four recent exclusive
semileptonic decays described above; (2) a recent calculation of $B
\to D^*\ell\nu$ \cite{Bailey:2014tva}; and (3) the result for the
ratio $|V_{cb}/V_{ub}|$ from a recent calculation of the exclusive
semileptonic decay ratios $\Gamma(\Lambda_b \to p \ell
\nu)/\Gamma(\Lambda_b \to \Lambda_c \ell \nu)$ by Detmold, Lehner, and
Meinel \cite{Detmold:2015aaa}.  The result of a simple weighted
average of all six lattice calculations gives $|V_{cb}| =
39.78(42)\times 10^{-3}$ and $|V_{ub}| = 3.59(9) \times 10^{-3}$.  The
nonlattice inclusive values from
Refs.~\cite{Alberti:2014yda,Amhis:2014hma} plotted there differ by
several standard deviations.

\subsection{Neutral $B$-meson mixing}

Mixing in the $B^0_x$ ($x=d,s$) system occurs at second order in the
electroweak interaction; it is parameterized by, among other terms,
the CKM matrix element $|V_{tx}|$ and the hadronic expectation value
$\langle O_{1x}\rangle$ of a $\Delta B = 2$ four-quark operator
$O_{1x}$.  The mixing strength can be parameterized by the
experimentally measured mass splitting $\Delta M_B$.  It is popular to
consider the ratio $\xi$ of mixing strengths for $B^0$ and $B^0_s$ for
which many theoretical uncertainties cancel
\begin{equation}
  \xi = \frac{M_{B^0}}{M_{B^0_s}}\sqrt{\frac{\langle O_{1s}\rangle}{\langle O_{1d}\rangle}}
      = \left|\frac{V_{td}}{V_{ts}}\right|\sqrt{\frac{\Delta M_{B^0_s}}{\Delta M_{B^0}} \frac{M_{B^0}}{M_{B^0_s}}} \,.
\end{equation}
Thus the combination of theory and experiment yields the ratio
of CKM matrix elements $|V_{td}/V_{ts}|$.

In the past year, a new lattice calculation \cite{SimoneLat15} has
produced a preliminary value $\xi = 1.210(19)$, which is compared with
recent results in Fig.~\ref{fig:excl_incl_xi}.  It yields a new, preliminary
value $|V_{td}/V_{ts}| = 0.2069(6)_{\rm exp}(32)_{\rm thy}$.  Here
experiment is still quite a bit ahead of theory in precision.

\begin{figure}
\centering
   \includegraphics[width=0.6\textwidth]{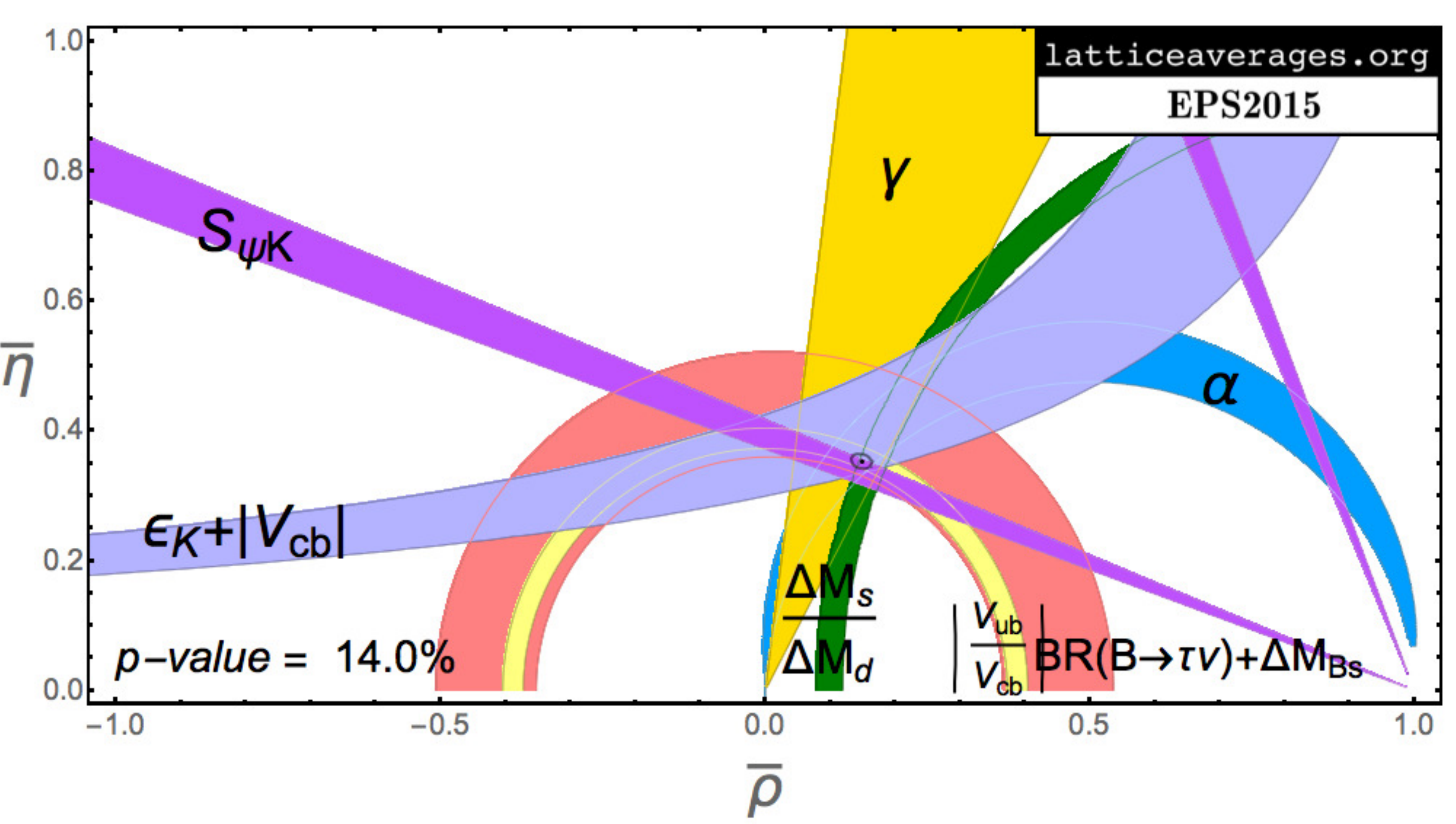}
  \caption{Unitarity triangle fit using the new average $|V_{ub}|$ and
    $|V_{cb}|$ exclusive values from
    Fig.~\protect\ref{fig:excl_incl_xi} and the new, preliminary value
    of $\xi$ from Ref.~\cite{SimoneLat15}. (Figure credit:
    \cite{Laiho:2009eu}).  }
 \label{fig:UTVcbVubxi}
\end{figure}

The ``unitarity triangle'' provides a graphical illustration of the
orthogonality of two rows of the CKM matrix as expected from
unitarity.  In Fig.~\ref{fig:UTVcbVubxi} we show the effect on the
unitarity triangle of recent results from exclusive semileptonic
decays for $|V_{cb}|$ and $|V_{ub}|$ and the preliminary result for
$|V_{td}/V_{ts}|$ discussed above.  So far the result is compatible
with three-generation CKM unitarity.

\subsection{Flavor-changing neutral currents}

\begin{figure}
\centering
\includegraphics[width=0.45\textwidth]{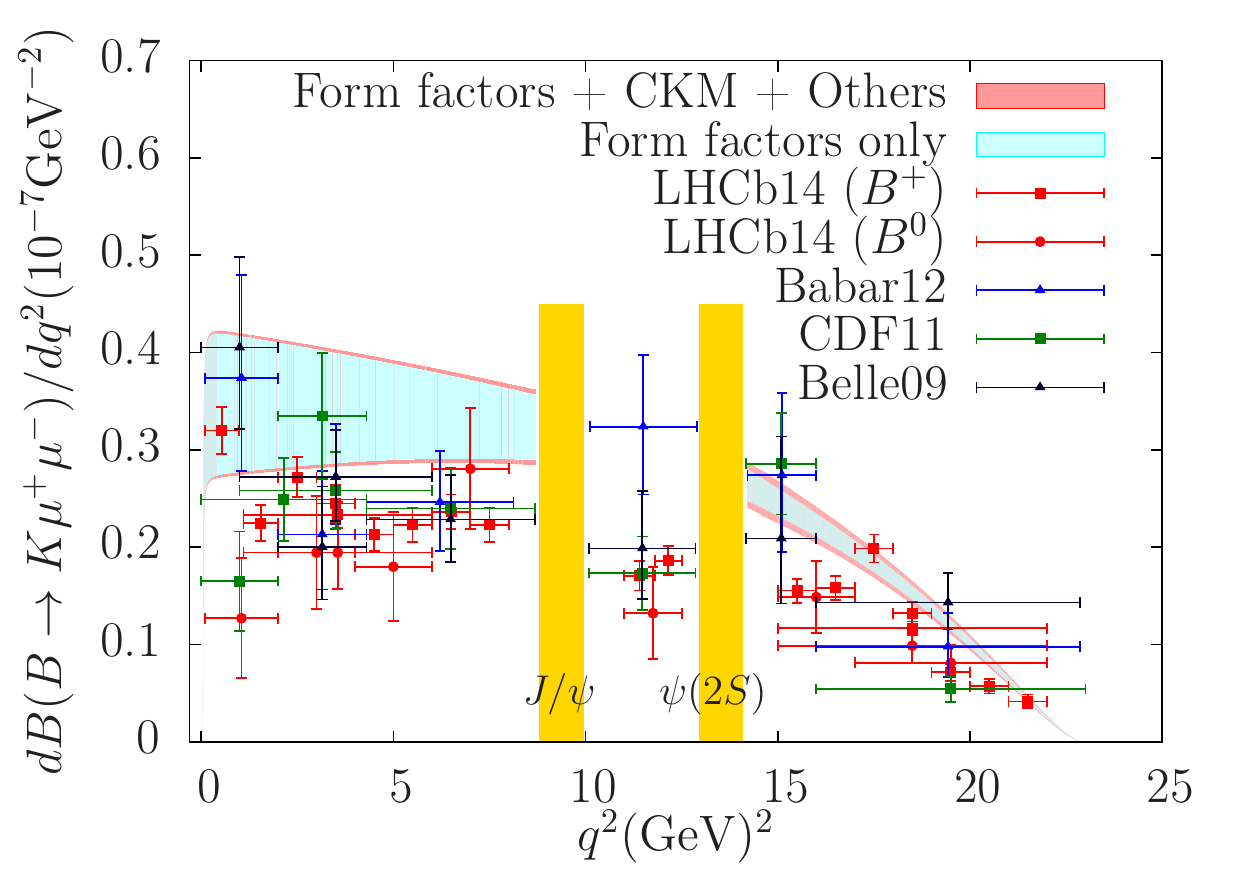} \hfill
\includegraphics[width=0.45\textwidth]{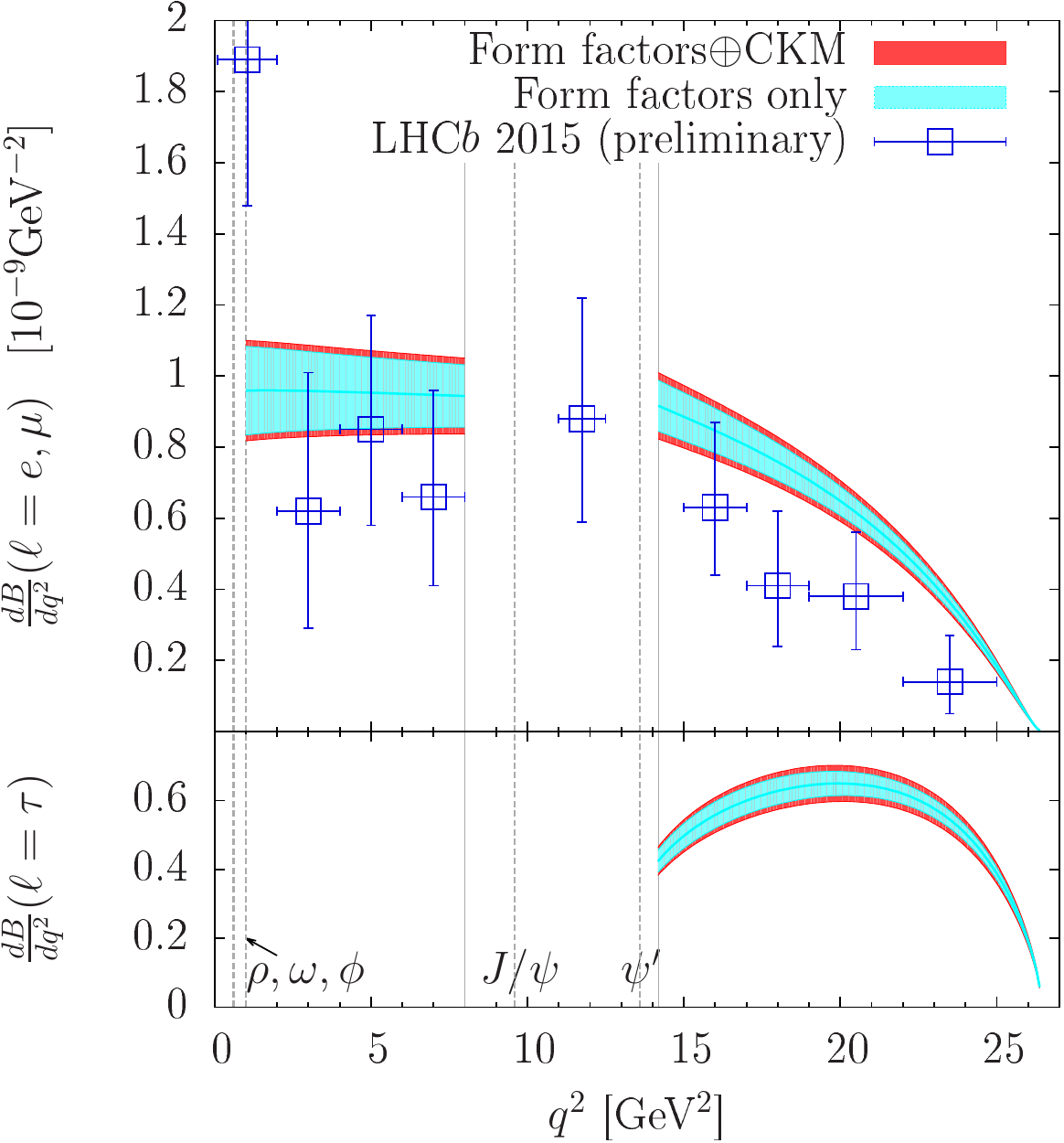}
    \caption{Left: A comparison of a recent standard-model lattice
      calculation of the differential branching fraction for $B^+\to
      K^+\mu^+\mu^-$ with experimental measurements
      \protect\cite{Wei:2009zv,Aaltonen:2011qs,Lees:2012tva,Aaij:2014pli}. The
      lattice values are from an improved analysis
      \protect\cite{Du:2015tda} of form factor results from
      \protect\cite{Bailey:2015dka}. The Belle, CDF, and BaBar
      experiments report isospin averages. Right: A similar comparison
      for the observable $B^+\to \pi^+\mu^+\mu^-$.  The lattice result
      is from \protect\cite{Bailey:2015nbd} and the experimental
      measurement is from LHCb \protect\cite{Aaij:2015nea}.  }
 \label{fig:BPSmumu}
\end{figure}

The processes $B \to \pi \ell\ell$ and $B \to K \ell\ell$ occur at
second-order in the electroweak interaction and are sensitive to new
flavor-changing-neutral-current processes.  Recent lattice
calculations have produced improved differential decay rates
\cite{Lattice:2015tia,Bailey:2015dka,Bailey:2015nbd}.  See also
Ref.~\cite{Bouchard:2013mia}. They are compared in
Fig.~\ref{fig:BPSmumu} with recent experimental measurements.  For
both processes the lattice values tend to lie slightly above the
experimental points.  The combined tension with the standard model is
1.7$\sigma$ \cite{Du:2015tda}.

\subsection{Direct CP violation in $K$ decays}

\begin{table}
  \caption{Comparison of results from the RBC-UKQCD collaboration
    \protect\cite{Bai:2015nea} and from Ishizuka {\it et al.}
    \cite{Ishizuka:2015oja} with experimental values
    \protect\cite{PDGCP} for the $I = 0$ and $I = 2$ amplitudes for
    the decay $K \to \pi\pi$ and for the direct CP-violating ratio
    $\epsilon^\prime/\epsilon$. The RBC-UKQCD calculation is done with
    physical kinematics.  Errors in the Ishizuka {\it et al.} values are
    statistical only.}
\label{tab:Kpipi}  
\begin{center}
\begin{tabular}{lccc}
       & RBC-UKQCD & Ishizuka {\it et al.} & experiment \\
 \hline
  $\Re(A_0) \times 10^8$ (GeV) & $46.6(10.0)_{\rm stat}(12.1)_{\rm sys}$ & $24.26(38)$ & $33.201(18)$ \\
  $\Re(A_2) \times 10^8$ (GeV) & $1.50(4)_{\rm stat}(14)_{\rm sys}$ & 60(36) & $1.474(4)$ \\
  $\Re(\epsilon^\prime/\epsilon)\times 10^4$ & $1.38(5.15)_{\rm stat}(4.43)_{\rm sys}$ & $0.8 (3.5)$ & $16.6(2.3)$ \\
\end{tabular}
\end{center}
\end{table}

Calculating the amplitudes for the purely hadronic weak decay $K \to
\pi\pi$ has been a decades-long challenge for lattice gauge theory.
In the past year the RBC-UKQCD collaboration announced the first
controlled lattice calculation with physical kinematics that accounts
numerically for the famous $\Delta I = 1/2$ rule -- namely, that the
amplitude for the isospin zero ($I = 0$) final state is considerably
larger than that of the $I = 2$ final state \cite{Bai:2015nea}.  The
calculation also yielded a value for the direct CP-violating ratio
$\epsilon^\prime/\epsilon$.  The calculation was done with domain-wall
fermions.  The new results are listed in Table~\ref{tab:Kpipi} and
compared with recent results from Ishizuka {\it et al.}
\cite{Ishizuka:2015oja} and experimental values.  The RBC-UKQCD result
for $\Re(\epsilon^\prime/\epsilon)$ is $2.1 \sigma$ below the
experimental value.

\subsection{Quark-flavor-physics conclusions}

A wealth of new data from KEK-$B$ and the LHC plus high-precision
lattice-QCD calculations are producing ever more stringent tests of
the standard model.  The tension between inclusive and exclusive
determinations of $|V_{cb}|$ and $|V_{ub}|$ remains.  New searches for
physics beyond the standard model appeared this year. There is some
tantalizing tension in rare $B$ decays, but still no unambiguous signs
of new physics.  We look forward to Belle-II, BES-III, and the LHC Run
2.  In the meantime, the lattice-QCD community must work hard to keep
up with experiment!

\section{Spectroscopy highlights}

Given the limitations of time, I will concentrate on calculations in
the charmonium sector where there has been good recent progress.  I
will first discuss how the lattice methodology has evolved.  For a
more extensive review, see Ref.~\cite{Prelovsek:2015fra}.

The time-honored method for determining the hadron spectrum works with
the correlator of a pair of hadron interpolating operators with
suitable quantum numbers, for example, the fermion bilinear ${\cal O}
= \bar q \gamma_\mu q$.  The correlation function between pairs
selected from a set of such operators,
\begin{equation}
  C_{ij}(t) = \langle 0|{\cal O}_i(t) {\cal O}_j(0)|0 \rangle \,,
\end{equation}
is a matrix. In terms of the eigenenergies $E_n$ of the hamiltonian,
it takes the multiexponential form
\begin{equation}
  C_{ij}(t) = \sum_n Z_{in} \exp(-E_nt) Z_{nj} \, .
\end{equation}
Since the lattice has a finite volume, the spectrum $E_n$ is always
discrete.  At sufficiently large time $t$ the rhs is dominated by the
lowest energy states. Those are the most reliably determined energies.
For recent charmonium examples, see
Refs.~\cite{Galloway:2014tta,Bali:2015lka}.

\begin{figure}
\centering
\includegraphics[width=0.45\textwidth]{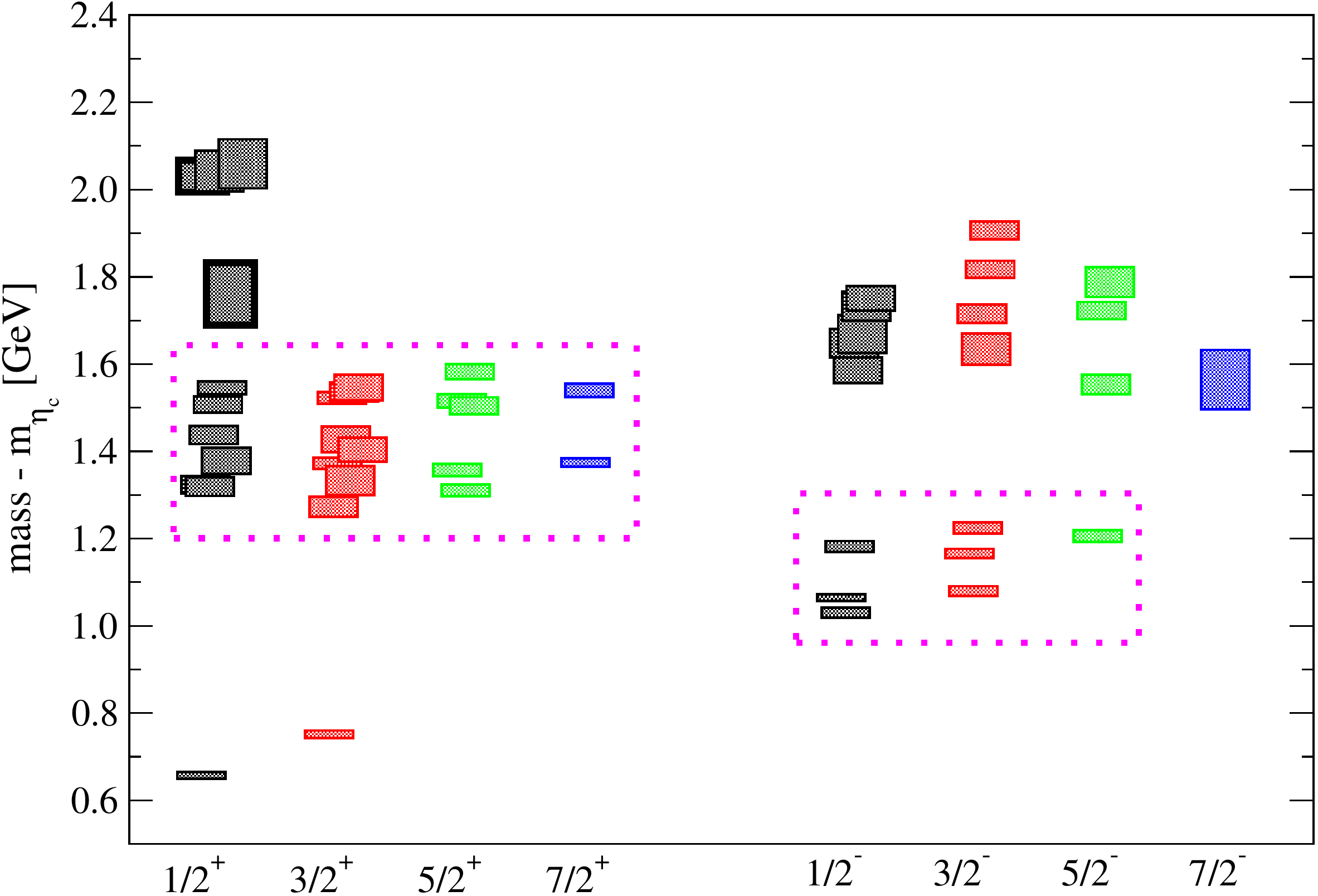} \hfill
\includegraphics[width=0.45\textwidth]{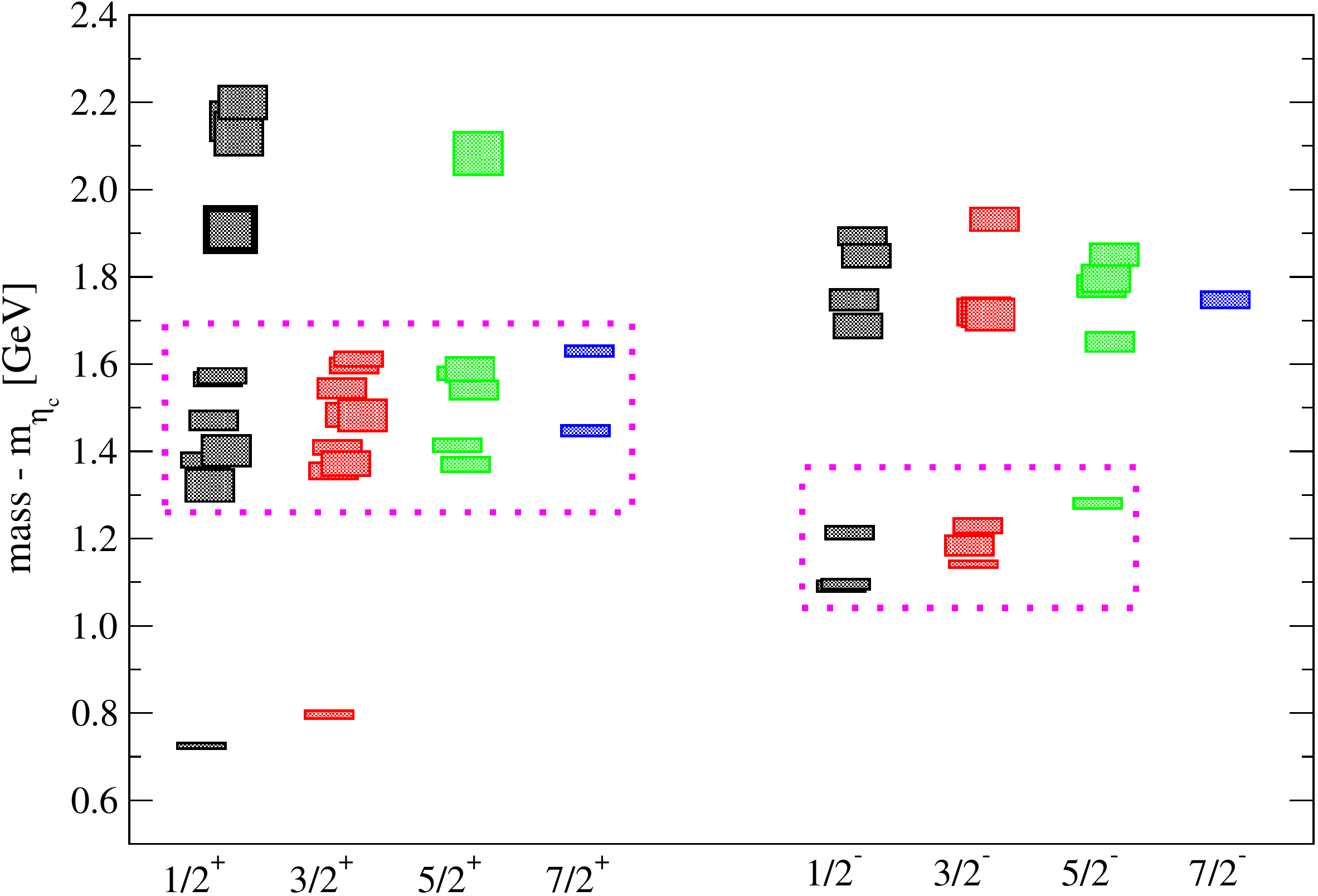}  
  \caption{Spectrum of charm $C = 2$ baryons (left panel, $\Xi_{cc}$';
    right panel, $\Omega_{cc}$) from a recent calculation by the
    Hadron Spectrum Collaboration \protect\cite{Padmanath:2015jea}.
\label{fig:C2baryons}}
\end{figure}

The problem of extracting the eigenenergies can be reformulated as a
variational calculation.  The interpolating operators act on the
vacuum to produce a variational basis set ${\cal O}_i |0\rangle$. That
set then evolves in Euclidean time.  As with any variational
calculation, success in determining the eigenenergies depends on how
well the evolved variational basis set overlaps with the eigenstates
themselves.  So if the basis set consists of only quark bilinears, we
might worry that we would have trouble describing states with a large
open-charm component, such as $D \bar D$.  Thus calculations with only
quark bilinears have done well describing charmonium levels below the
open charm threshold, but become increasingly unreliable at higher
energies.  For an example of an impressive calculation of this type
for charmonium with many proposed excited and exotic charmonium
states, see Ref.~\cite{Liu:2012ze}.  In the past year the same methods
have been applied to charm $C = 2$ baryons as shown in
Fig.~\ref{fig:C2baryons}.

\begin{figure}
\centering
   \includegraphics[width=0.6\textwidth]{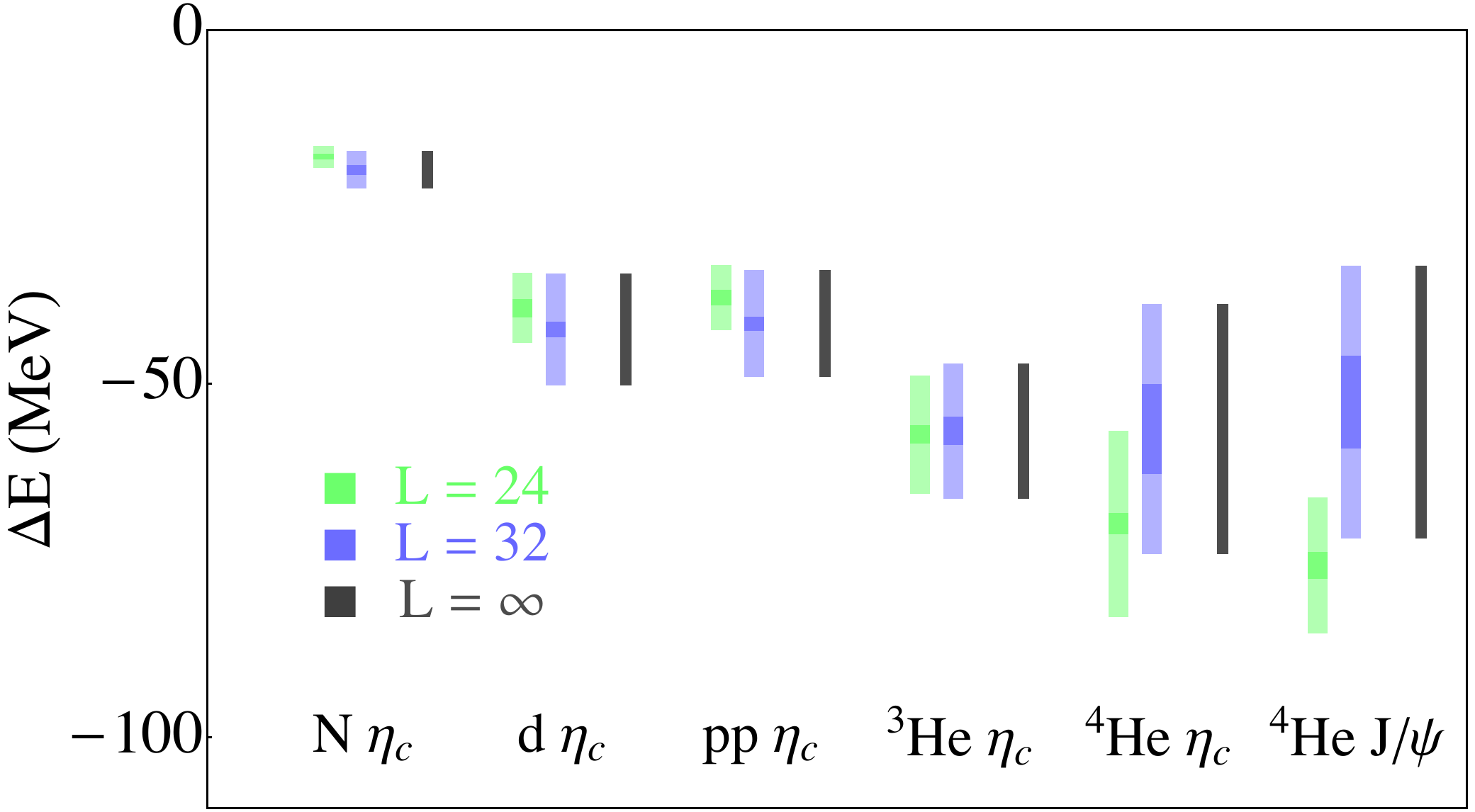}
  \caption{Binding energies for charmonium with light nuclei from
    Ref.~\cite{Beane:2014sda}, shown for two box sizes and an infinite
    volume extrapolation.  Caution: the calculation was done at the
    SU(3)-flavor-symmetric point, so with $M_\pi \approx 805$ MeV.}
 \label{fig:charmNbinding}
\end{figure}

Particularly to study states that couple strongly to nearby
multihadronic states, we must incorporate multihadronic states
explicitly in the analysis. For identifying weakly bound states, a the
more conventional nonvariational analysis suffices. An interesting
recent example, particularly in light of increasing evidence for
pentaquark states involving charm quarks \cite{Aaij:2015tga} is a
study of the binding of charmonium to nuclei by the NPLQCD
collaboration \cite{Beane:2014sda}. Here one starts with an
interpolating operator that creates both the charmonium state and the
nucleus in question and measures the difference between the ground
state energy of the combined, interacting system and the total energy
of the noninteracting component hadrons.  Some results are shown in
Fig.~\ref{fig:charmNbinding} and suggest binding energies of the order
of several tens of MeV for light nuclei.

For charmonium, in the past few years attention has shifted to the
study of excited levels using the variational method with two types of
interpolating operators: quark bilinears and multiquark operators that
create explicit open charm state, such as $D \bar D$.  Then it becomes
possible to study resonant and other states whose existence is
strongly influenced by nearby open charm levels.  For states below
inelastic threshold a method developed by L\"uscher leads to the
scattering phase shift.  In the case of a resonance one then
determines not only the resonant energy but also the width.  In the
case of a shallow bound state, one can also determine the binding
energy.
n
\begin{figure}
\centering
   \includegraphics[width=0.7\textwidth]{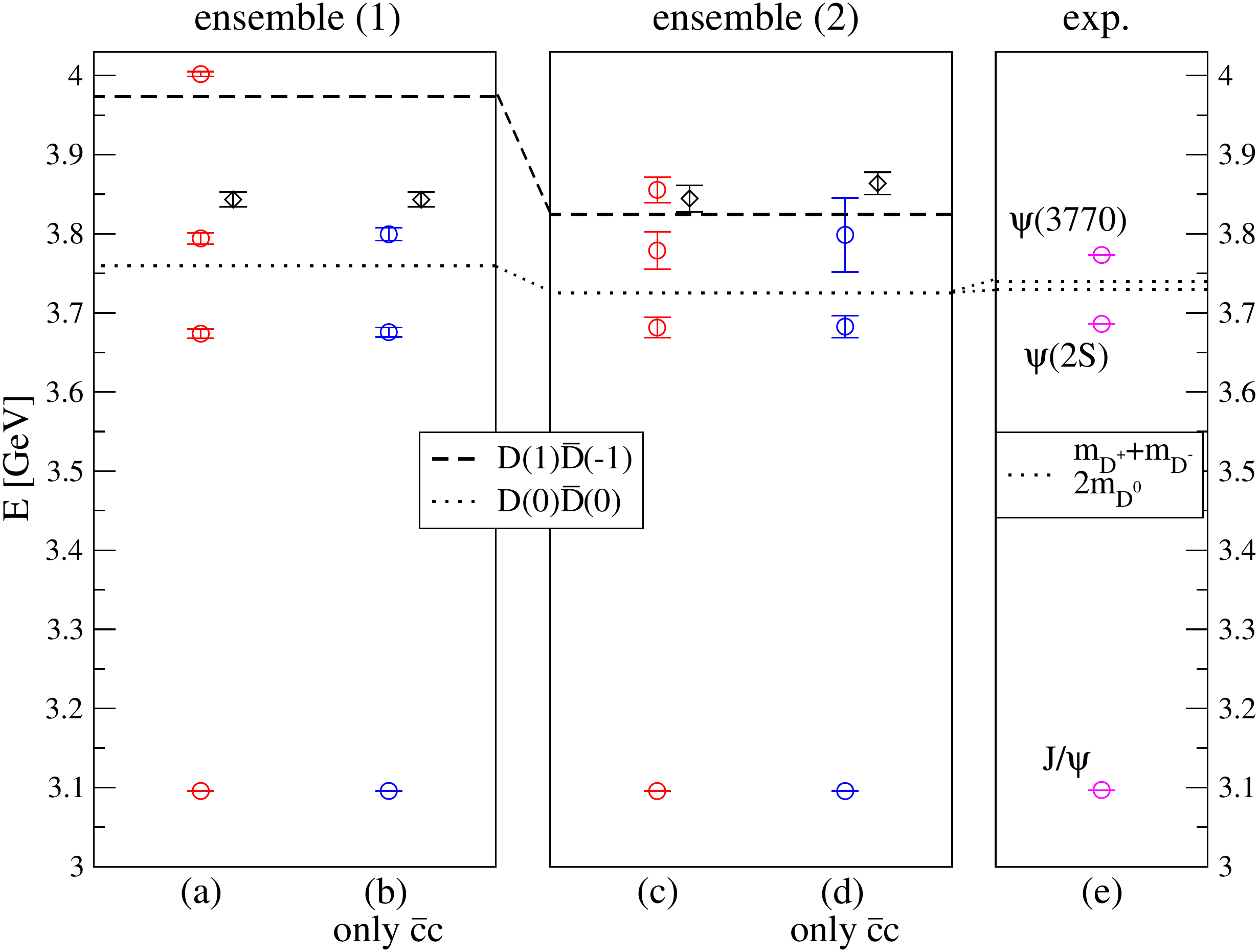}
  \caption{Finite box energies for the charmonium $T_1^{--}$ channel
    containing both $1^{--}$ and $3^{--}$ states from
    Ref.~\protect\cite{Lang:2015sba}. Two different box sizes are used (left: $L =
    1.98$ fm, middle: $L = 2.90$ fm).  Dashed lines show
    noninteracting energies of two discrete $D(p) \bar D(-p)$
    scattering states with $p = 2\pi/L$ and $4\pi/L$.
    Red points come from a full variational basis including
    both $\bar c c$ and $D(p) \bar D(-p)$ state.  Blue points
    come from a reduced basis restricted to $\bar c c$.  Black
    diamonds indicate a $3^{--}$ assignment.
}
 \label{fig:cc3770}
\end{figure}

The L\"uscher method exploits the physics of interacting hadrons
confined to a finite box \cite{Luscher:1991cf}.  In a finite box the
spectrum is discrete and depends on the linear box dimension $L$.  If
one can assume that the interaction takes place over a distance $R$
less than $L/2$ then the discrete box energies carry information about
the infinite-volume elastic scattering amplitude --- hence the
infinite-volume elastic phase shift. In the simplest case of two
identical particles of mass $M$, one uses
\begin{equation}
   E_n = 2\sqrt{p_n^2 + M^2}
\end{equation}
to convert a given discrete box energy level $E_n$ into an
``interacting momentum'' $p_n$.  The L\"uscher analysis then leads to
an infinite-volume phase shift $\delta_n$ at that momentum
$p_n$. These values can be interpolated to obtain $\delta(p)$.  At a
different value $L$ one can obtain a new set of values for the same
function $\delta(p)$ to improve the interpolation.  In the $S$ wave,
the infinite-volume elastic scattering amplitude is, as usual,
\begin{equation}
  T(p) = \frac{1}{p \cot \delta(p) - ip} \, .
\end{equation}
Poles correspond to bound states or resonances.  The method can be
generalized to the inelastic (coupled-channel) case
\cite{Hansen:2012tf,Briceno:2012yi,Guo:2012hv}.

\begin{figure}
\centering
\includegraphics[width=0.45\textwidth]{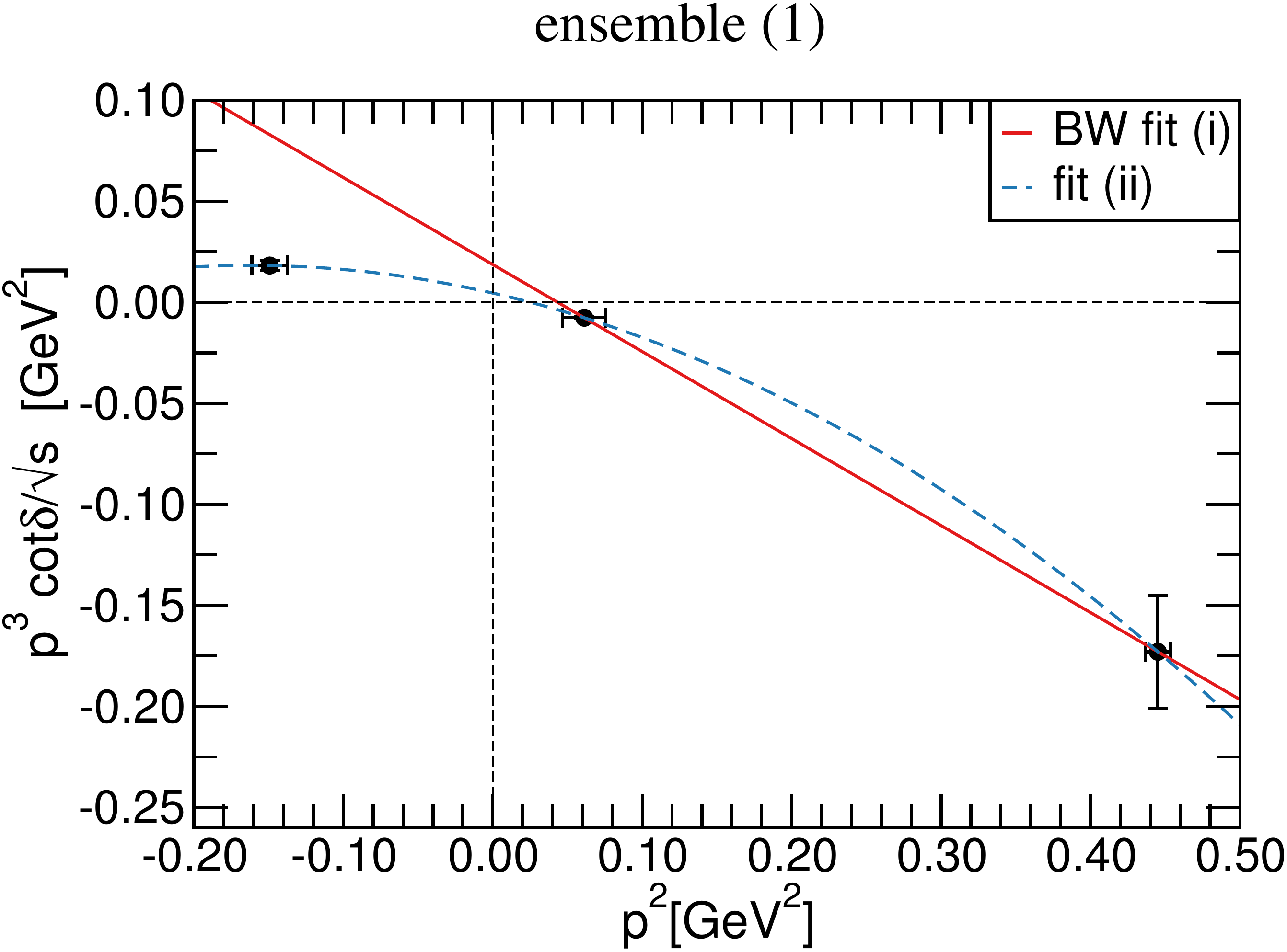}
\hfill
\includegraphics[width=0.45\textwidth]{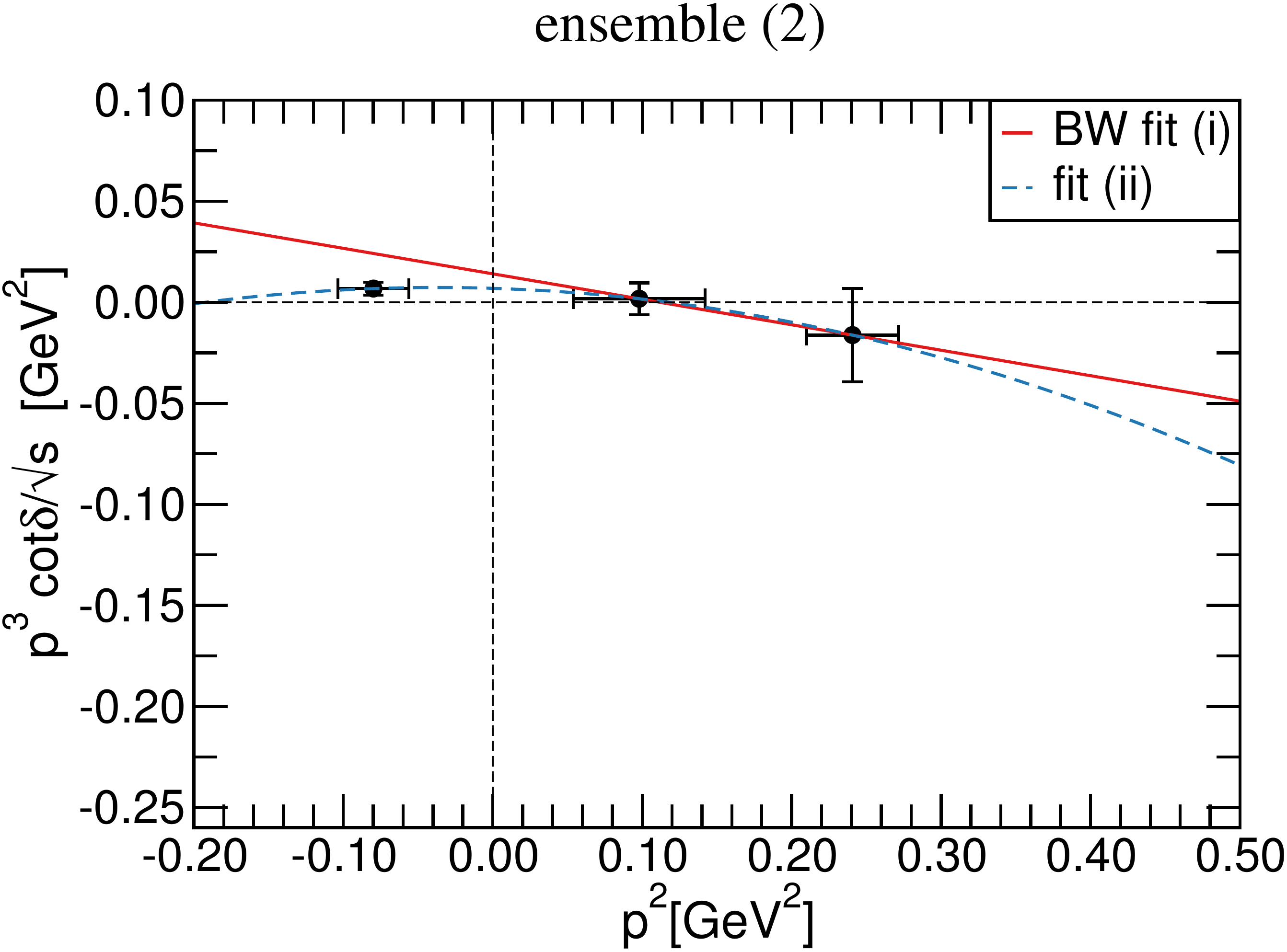}
  \caption{Phase shifts resulting from the finite-box states in
    Fig.~\protect\ref{fig:cc3770} from
    Ref.~\protect\cite{Lang:2015sba}.  }
\label{fig:cc3770phase}
\end{figure}

When used in conjunction with multiple interpolating operators, the
L\"uscher method becomes a powerful tool for studying excited
charmonium states.  For a recent example, we point to work by Lang,
Lescovec, Mohler, and Prelovsek \cite{Lang:2015sba} who looked at the
charmonium state $\psi(3770)$, a $J^{PC} = 1^{--}$ $P$-wave resonance
just above the $D \bar D$ threshold.  Figure~\ref{fig:cc3770} shows
their discrete box energies for two different box sizes and different
sets of interpolating operators.  The resulting phase shifts are
plotted in Fig.~\ref{fig:cc3770phase}.  The zero in $p \cot \delta(p)$
just above $DD$ threshold is identified as the $\psi(3770)$.

\begin{figure}
\centering
  \includegraphics[height=3.6cm,clip]{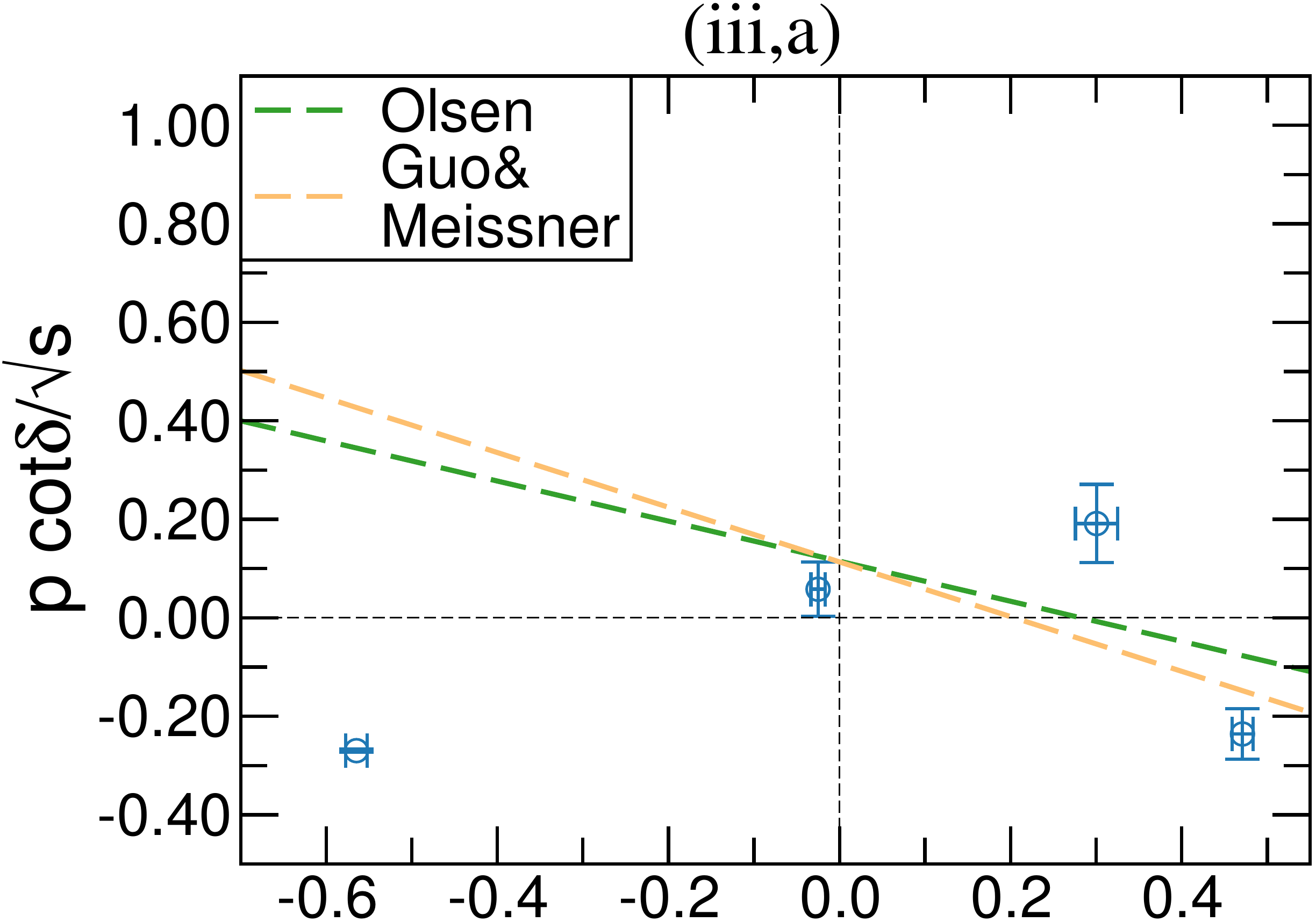}
  \includegraphics[height=3.6cm,clip]{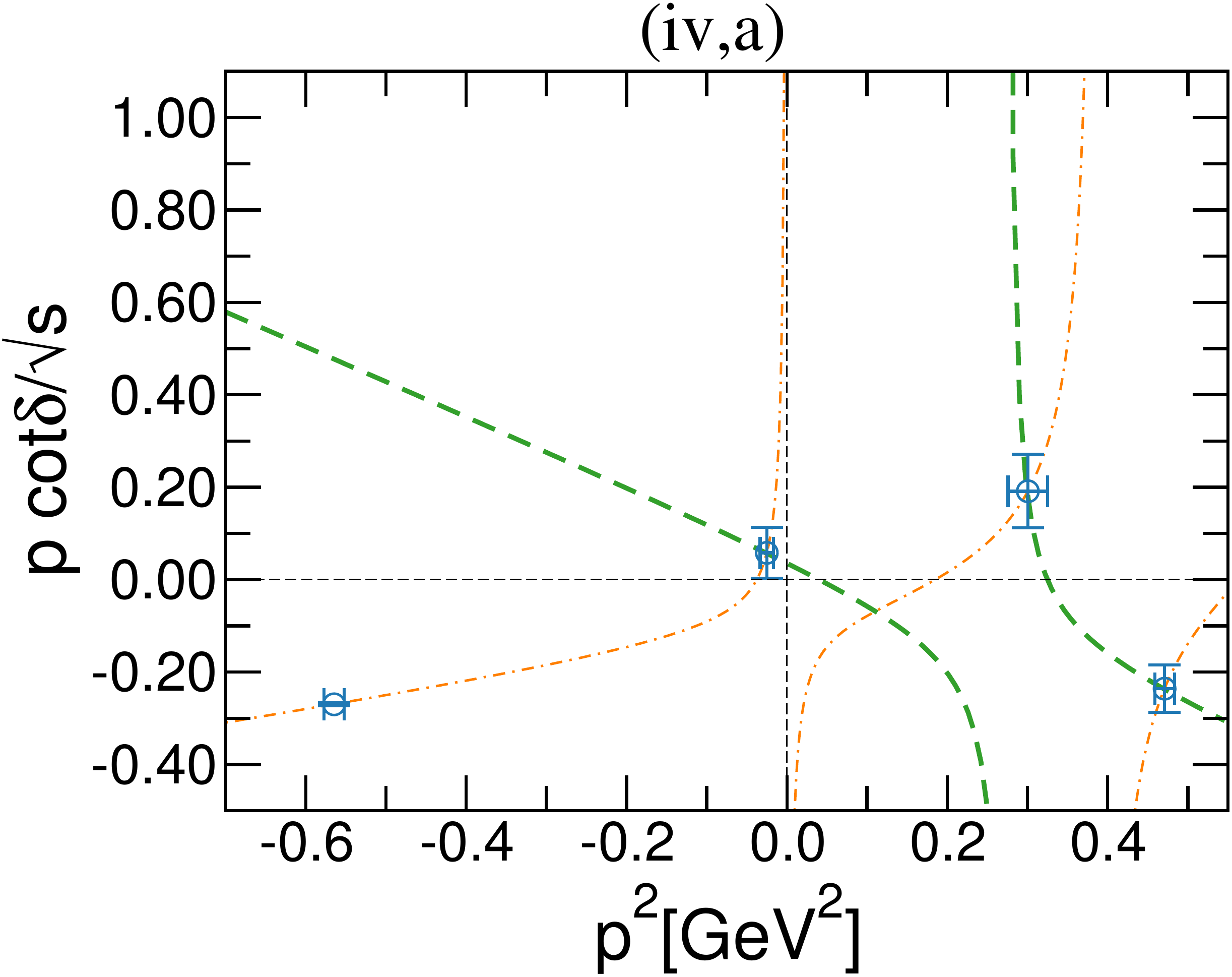}
  \includegraphics[height=3.6cm,clip]{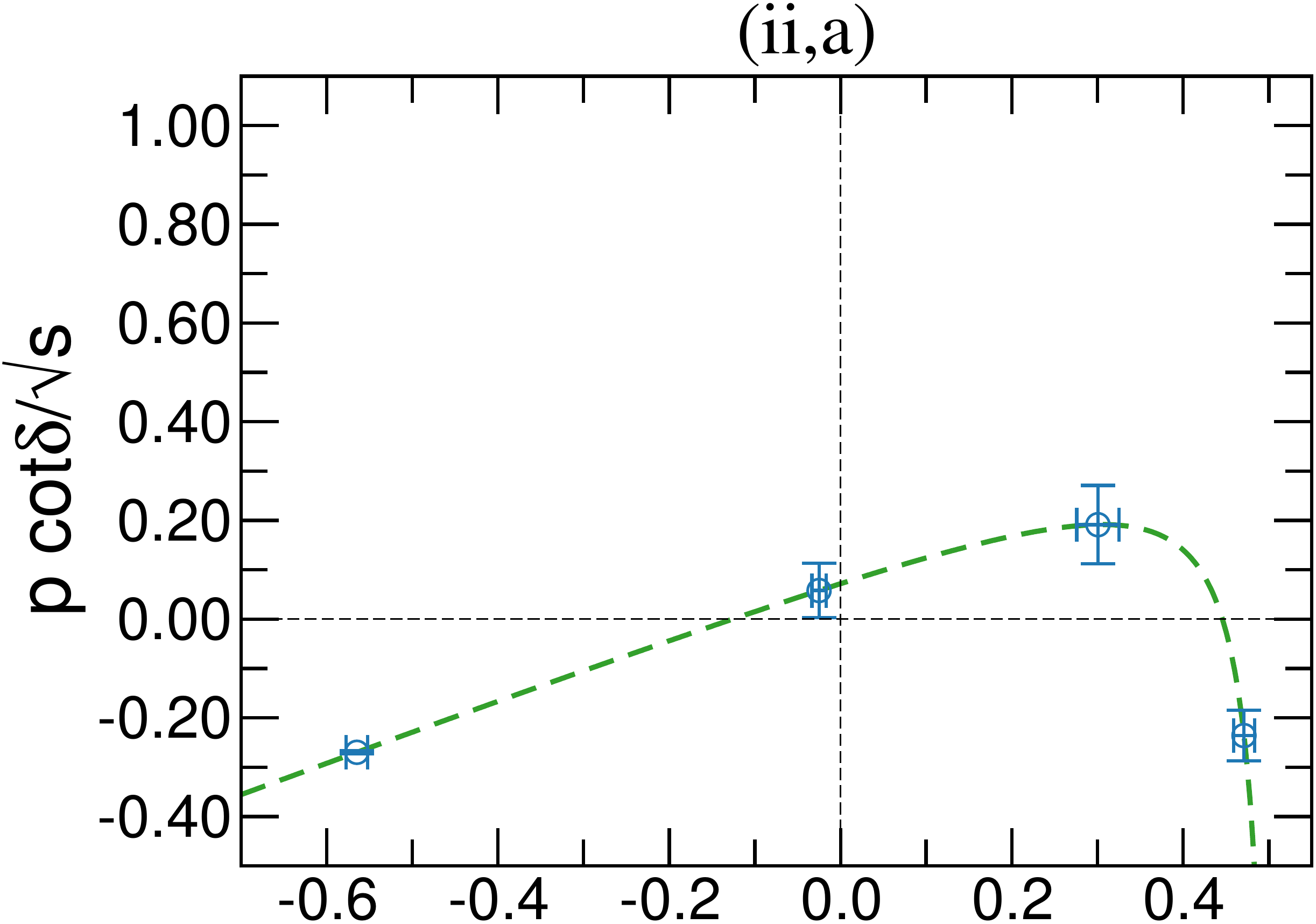}
  \caption{Results of fitting the lattice scalar $D \bar D$ phase
    shift to a variety of parameterizations in a study of the
    $X(3915)$ \cite{Lang:2015sba}.  Left (rejected): one broad
    Breit-Wigner resonance, middle (rejected): two Breit-Wigner
    resonances, right (preferred): one narrow resonance and the
    $\chi_{c0}(1P)$ state. }
 \label{fig:cc3915}
\end{figure}

In the same project, Lang, Lescovec, Mohler, and Prelovsek also looked
at the $X(3915)$, judged by the PDG to be a narrow scalar
enhancement above threshold, perhaps the charmonium $\chi_{c0}(2P)$
state. Lang {\it et al.} extract the phase shift and try a variety of
parameterizations. Results for various hypotheses are shown in
Fig.~\ref{fig:cc3915}. Their preferred parameterization is consistent
with the well-known $1P$ bound state plus a narrow resonance around
approximately 4 GeV.  However, Zhou, Xiao, and Zhou make a case that
the correct $J^{PC}$ assignment for the state is $2^{++}$
\cite{Zhou:2015uva}, so this story is still unfinished.

\begin{figure}
\centering
   \includegraphics[width=0.6\textwidth]{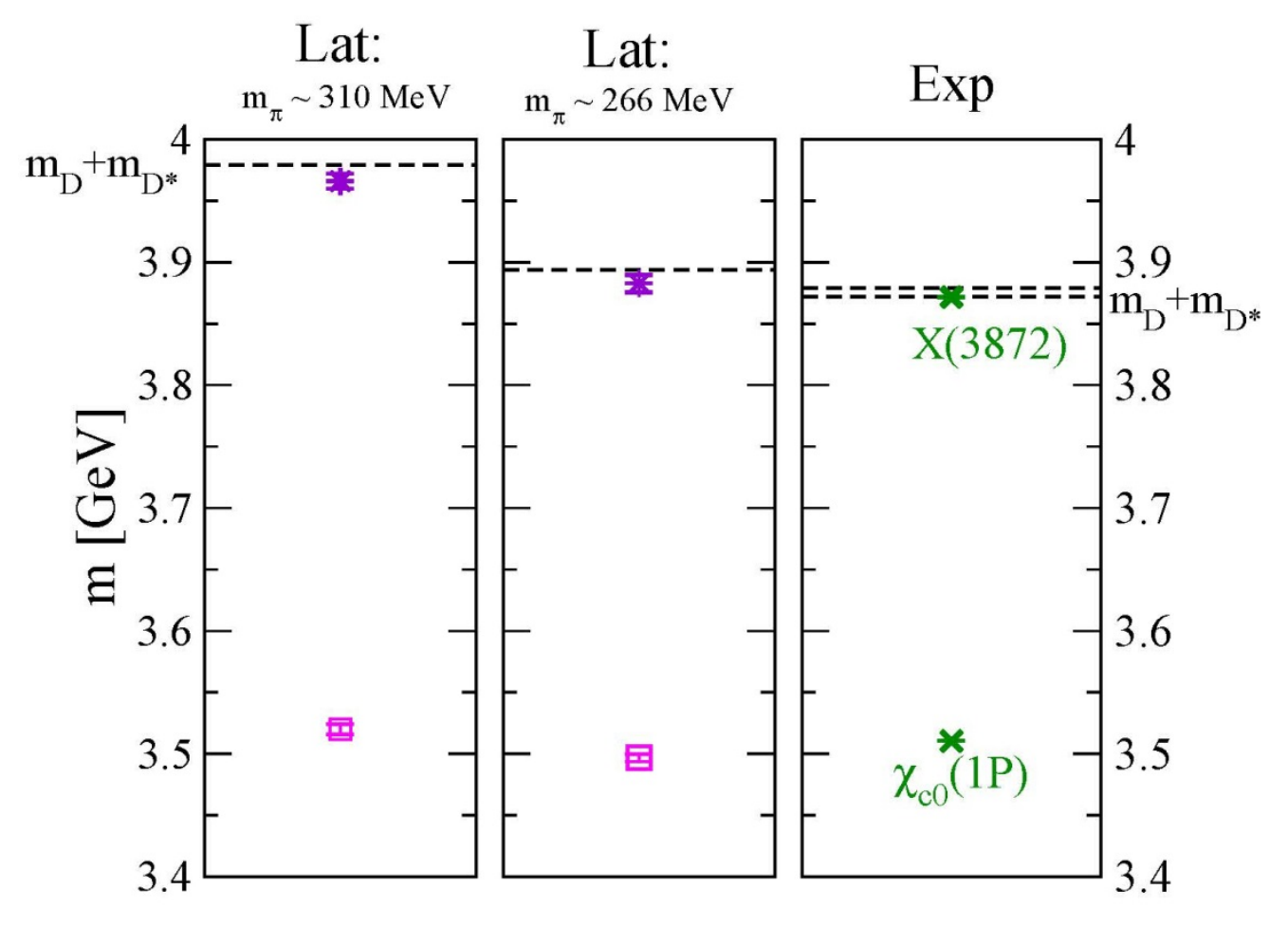}
  \caption{Comparison of results for the spectrum in the charmonium
    $1^{++}$ channel from two lattice calculations. Left:
    Ref.~\cite{Lee:2014uta}, middle: Ref.~\cite{Prelovsek:2013cra}
    and right: experimentally measured masses. Dashed lines indicate
    the $D \overline{D^*}$ threshold. The $X(3872)$ candidate is
    just below it. (Figure credit: \cite{Prelovsek:2015fra}.}
 \label{fig:X3872}
\end{figure}

The $1^{++}$ charmonium state $X(3872)$ has been the subject of study
in lattice QCD for the past few years. It is remarkably close to (less
than 1 MeV below) the $D \overline{D^*}$ threshold.  It is therefore
expected to be strongly influenced by the open charm threshold.  The
quark model predicts a $\chi_{c1}(2P)$ state in the vicinity.  For
these reasons the theoretical understanding of the state should take
into account both $\bar c c$ states and open charm state.  Results for
the spectrum from calculations by Lescovec and Prelovsek
\cite{Prelovsek:2013cra} and later calculations by Lee {\it et al.}
\cite{Lee:2014uta} are shown in Fig.~\ref{fig:X3872}.  These
calculations are done with different lattice ensembles, different
light valence quarks at unphysical masses and at only one lattice
spacing.  However, both see an $X(3872)$ candidate as a shallow bound
state of $D \overline{D^*}$. Both use the L\"uscher method, but some
phenomenological arguments suggest that the state is quite large
(perhaps 6 fm) \cite{Braaten:2009zz}, larger than the lattice box
size, so it is important to test the result for finite volume
effects.

\begin{figure}
\centering
   \includegraphics[width=0.45\textwidth]{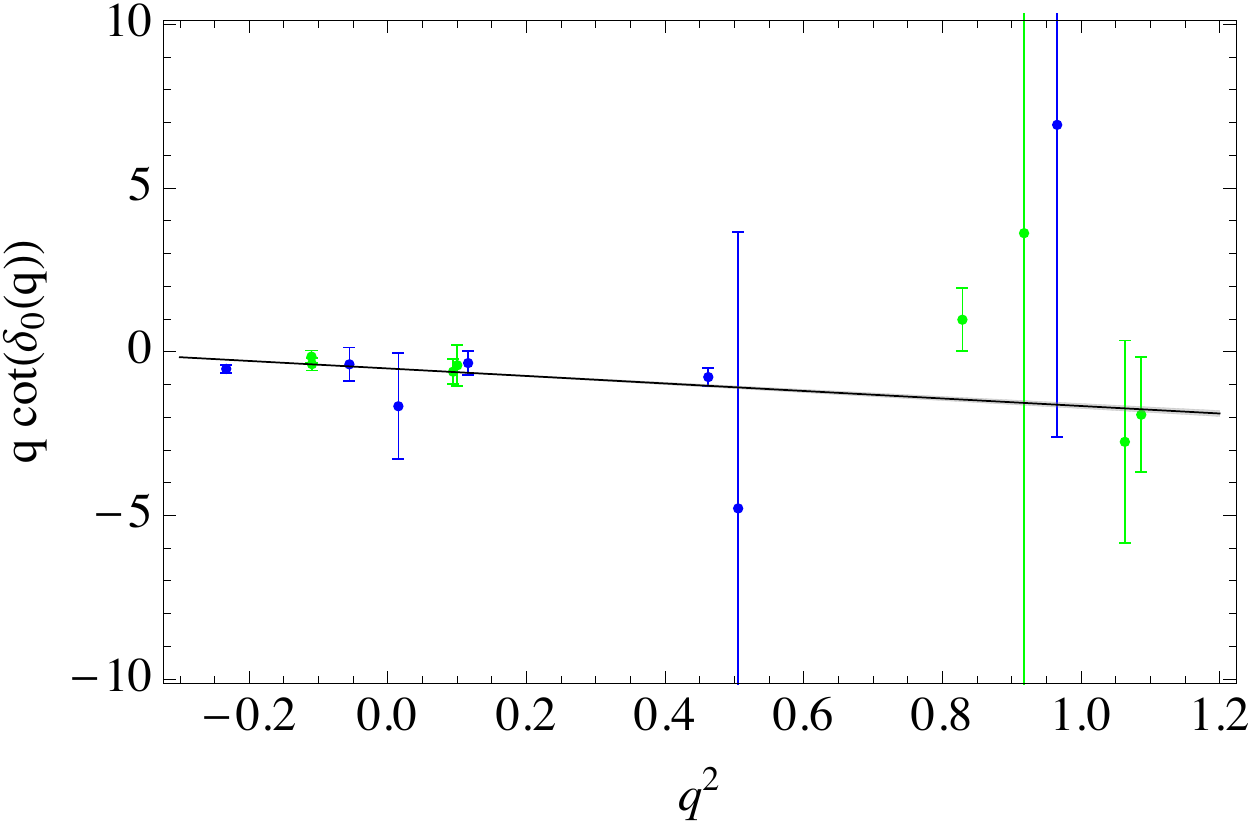} \hfill
   \includegraphics[width=0.45\textwidth]{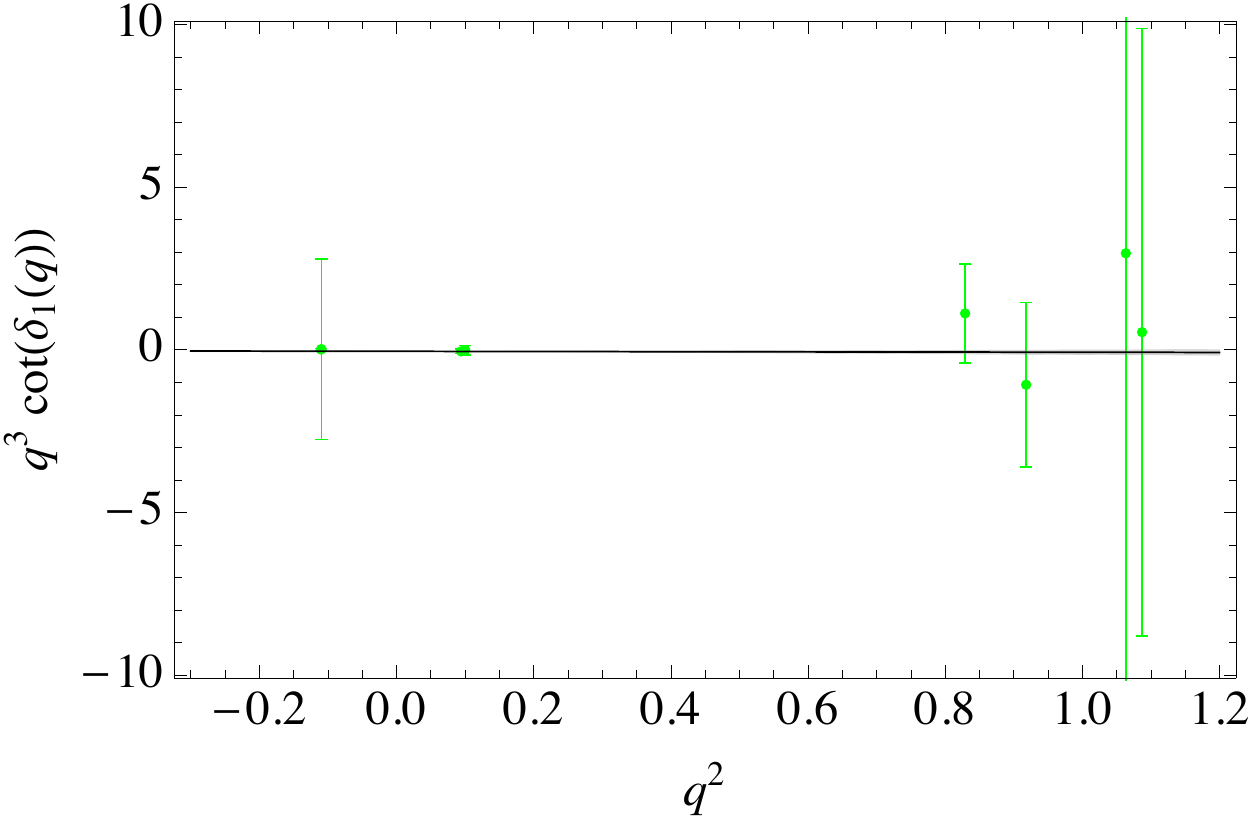}
  \caption{The $S$-wave and $P$-wave phase shifts for $D \bar D^*$
    scattering from a recent lattice QCD study by the CLQCD
    collaboration \protect\cite{Chen:2014afa}.  There is no sign of a
    resonance in either channel.
 \label{fig:Zc3900}
}
\end{figure}

Unlike the $X(3872)$, the $Z_c(3900)$ has not been found in lattice
QCD studies so far \cite{Chen:2014afa}.  The result of a recent effort
is shown in Fig.~\ref{fig:Zc3900}. No $S$- or $P$-wave resonance was
found.

\subsection{Spectroscopy conclusions}

New spectroscopic methods combine a multichannel analysis with
L\"uscher's formalism.  There has been rapid progress in understanding
charmonium resonances and bound states close to an elastic threshold.
The $X(3872)$ could be a shallow $D\overline{D^*}$ bound state but
more work is needed to check finite size effects. The X(3915) needs
more study. The $Z_c(3900)$ has thus far escaped detection in a
lattice study.

\paragraph{Acknowledgments} I am grateful to Andreas Kronfeld, Daniel Mohler, and Ruth Van de Water for
assistance with these proceedings.  This work is supported by the U.S.
National Science Foundation under grant PHY10-034278.

\end{document}